\documentclass[prb,superscriptaddress, showpacs,twocolumn,papersize=a4paper]{revtex4}

\usepackage[unicode]{hyperref}
\usepackage{amsmath}
\usepackage{amsfonts}
\usepackage{amssymb}
\usepackage{graphicx}
\usepackage{color}
\usepackage{bm}

\newcommand{\be}{\begin{eqnarray}}
\newcommand{\ee}{\end{eqnarray}}
\newcommand{\ba}{\begin{array}}
\newcommand{\ea}{\end{array}}
\newcommand{\no}{\nonumber}

\newcommand{\sign}{\mathop{\rm sign}\nolimits}

\begin{document}
\title{Fluctuation conductivity in disordered superconducting films}
\author{Konstantin S. Tikhonov}
\affiliation{Department of Physics, Texas A\&M University, College Station, TX
77843-4242, USA}
\affiliation{L. D. Landau Institute for Theoretical Physics, 117940 Moscow, Russia}
\email{tikhonov@itp.ac.ru}
\author{Georg Schwiete}
\affiliation{Dahlem Center for Complex Quantum Systems and Institut f\"ur Theoretische
Physik, Freie Universit\"at Berlin, 14195 Berlin, Germany}
\author{Alexander M. Finkel'stein}\affiliation{Department of Physics, Texas A\&M University, College Station, TX
77843-4242, USA}
\affiliation{Department of Condensed Matter Physics, The Weizmann Institute of Science,
76100 Rehovot, Israel}
\date{\today}
\begin{abstract}
We study the effect of superconducting fluctuations on the longitudinal and the transverse (Hall) conductivity in homogeneously disordered films. Our calculation is based on the Usadel equation in the real-time formulation. We adjust this approach to derive analytic expressions for the fluctuation corrections in the entire metallic part of the temperature-magnetic field phase diagram, including the effects of both classical and quantum fluctuations. This method allows to obtain fluctuation corrections in a compact and effective way, establishing a direct connection between phenomenological and microscopic calculations.
\end{abstract}

\maketitle

\section{Introduction}

Theoretical studies of fluctuation conductivity in superconductors found
their origin in the discovery of paraconductivity by Aslamazov and Larkin
(AL) in 1968 \cite{aslamazov68}. These authors analyzed the conductivity of
superconductors in the metallic phase above the transition temperature $%
T_{c} $ in the framework of diagrammatic linear response
theory. Paraconductivity can be understood as the direct contribution of
fluctuating Cooper pairs to the electric current. Indeed, the
formation of Cooper pairs opens a new channel for charge transport in the
metallic phase. Above the transition temperature, these Cooper pairs do not
form a condensate yet and their contribution to conductivity is positive but still bounded due to their finite lifetime.
Other effects of superconducting fluctuations are Andreev scattering of
electrons off the fluctuating order parameter described by the so-called Maki-Thompson (MT) term \cite{maki68,
thompson70}, and the suppression of the quasiparticle density of states (DOS) near the Fermi-level.

These classical results were obtained for temperatures close to $T_{c}$ and
later extended for larger temperatures and for weak
magnetic fields. More recently, the vicinity of the magnetic field-tuned quantum phase transition in disordered superconducting films was studied in a paper by Galitski and Larkin \cite{galitski01}. These
authors have shown that close to the quantum transition, contrary to the previously studied regime of weak magnetic fields, different processes are of equal importance. This has the
remarkable consequence that the sign of the total correction to conductivity becomes negative for sufficiently low temperatures near the quantum critical point, resulting in a non-monotonic magnetoresistance in this regime.

In spite of the substantial amount of existing theoretical work on
superconducting fluctuations, summarized in the book by Larkin and Varlamov \cite{larkin05}, the
subject continues to be an active field of research. This activity is stimulated by recent accurate experimental studies of different superconducting systems \cite{sanquer04,kapitulnik05,baturina05,behnia06,moler07,liu11}, that call for refined theoretical studies. For example, when
fitting experimental data on disordered superconducting films by theoretical
results, one commonly uses several fitting parameters,
including the critical temperature $T_{c}$, the upper critical field $B_{c}$
and the dephasing time $\tau _{\phi }$. In doing so, it would be useful to
work with theoretical results which are valid in the entire $(B,~T)$
phase diagram, instead of addressing different asymptotic regions separately. This is the motivation for the detailed calculations presented in this paper.

In deriving the results for the fluctuation conductivity, we deviate from
the traditional route that employs the diagrammatic linear response theory
in the imaginary time technique \cite{abricosov63} as described in detail, for
example, in Ref.~\onlinecite{larkin05}. Instead, we develop a formalism based on the Keldysh (real-time) representation of the Usadel equation. In this approach, disorder averaging is performed at the earliest
stages, thereby avoiding the use of the impurity-diagram technique. As an
additional advantage, no analytic continuation is required. The Usadel
equation \cite{usadel70} is an indispensable tool in the theory of mesoscopic
superconductors and hybrid structures \cite{belzig99,smith86}. This equation
describes low-energy (diffusive) physics on spatial $q^{-1}$ and temporal $%
\omega^{-1}$ scales, satisfying $\left( q l,~\omega \tau \right) \ll 1$%
, where $\tau$ is the impurity scattering time and $l$ the mean free path. The first calculation of superconducting fluctuation corrections in this framework was performed by Volkov et al. \cite{volkov98}, who calculated fluctuation conductivity in hybrid superconducting-normal structures in the vicinity of $T_{c}$ in the absence of a magnetic field.

In this paper, we use the Usadel equation to calculate longitudinal and transverse (Hall) conductivity in disordered superconducting films at arbitrary temperatures and magnetic fields. Our approach parallels to some extent the non-linear $\sigma$-model formalism for disordered superconductors introduced by Feigelman et al.~\cite
{feigelman00}, and the subsequent work by Kamenev and Levchenko \cite{levchenko07}. The latter work includes a calculation of fluctuation conductivity close to $T_c$. The intimate relation between the $\sigma$-model formalism and the Usadel equation approach is based on the fact that the Usadel equation is the saddle point equation of the nonlinear $\sigma$-model. For the sake of simplicity, we decided not to use the more technical apparatus of the nonlinear $\sigma$-model, but formulate the derivation in terms of the Usadel equation. This route leads us to a description in terms of a coupled set of kinetic equations for quasiparticles moving on the background of superconducting fluctuations. This method appears to be a  very convenient tool for studying fluctuation transport.

The classification of the fluctuation corrections obtained
in the discussed method appears to be very different from the conventional classification based on the diagrams in the
Matsubara technique. Therefore comparison with the results
obtained by the diagrammatic technique can be performed
only on the level of the final results. Let us mention here the comparison to recent works. It can be seen\cite{Tarasinski} that the
zero magnetic field limit of the general formulas
derived in this manuscript (Eqs. (\ref{DOSFinal}), (\ref{MTFinal}), and
(\ref{ALFinal}) below) can be presented in a form that exactly coincides with the corresponding diagrammatic results of Lopatin et al. in Refs. \onlinecite{Lopatin05,Shah07}.
On the other hand, Glatz et al. more recently presented a diagrammatic analysis of the
longitudinal fluctuation conductivity in the entire phase diagram \cite{glatz11,glatz11long}. However, their results are inconsistent with previous diagrammatic calculations as well as with ours (we comment on this work below at the end of Sec.~\ref{sec:results_long}).
For the Hall effect, our results agree with those of a work \cite{michaeli12} in which an independent calculation has been performed. These results were successfully applied for the description of a recent measurement in amorphous Tantalum Nitride films. \cite{breznay10}

This paper is organized as follows. In Sec.~\ref{sec:basic} we present the basic formalism. We show how the Usadel equation, initially formulated for a given order parameter configuration\cite{usadel70}, can be applied to the calculation of fluctuation conductivity. As a next step, in Sec.~\ref{sec:solution} we find a solution of the Usadel equation which allows to determine the order parameter correlation function in the Gaussian approximation. Both ingredients are required for the calculation of the electric current presented in Sec.~\ref{sec:calculation}. Next, we derive expressions for the longitudinal conductivity that are valid in the entire metallic phase outside the regime of strong fluctuations. Evaluation of the obtained expressions still requires a summation over the Landau levels as well as an integration over slow (bosonic) frequencies, which can be performed analytically only in certain limiting cases. Several such limiting cases are analyzed in detail in Sec.~\ref{sec:results_long}, including the region close to $T_c$ and the vicinity of the quantum critical point. By means of a numerical evaluation, we locate the line of the sign change for magnetoresistance $\partial\sigma/\partial B$ and the line $\partial\sigma/\partial T=0$. We also discuss the existence of a crossing point of the magnetoresistance curves. In Sec.~ \ref{sec:hall} we calculate Hall conductivity, generalizing previous calculations \cite{fukuyama71,aronov92,aronov95} to the case of arbitrary temperatures and magnetic fields above the transition.

\section{Basic equations}
\label{sec:basic}

In this section we present the equations that form the basis for our
calculation of the fluctuation conductivity. After stating the microscopic
model, we introduce the Usadel equation that allows to find the
quasiclassical Green's function in the dirty limit, i.e., if the condition $%
T_{c}\tau \ll 1$ is fulfilled. Calculation of the conductivity requires
knowledge of both the quasiclassical Green's function in the presence of the fluctuating order parameter field and the correlation
function of the order parameter field. In the fluctuation regime, which we study in this paper, the order parameter correlation function is governed by
the Ginzburg-Landau (GL) action. Fortunately, the GL action can be
found from the quasiclassical Green's function itself, i.e., from the
solution of the Usadel equation. This procedure will also be described
in this section.

We start with the Keldysh action for electrons with short-range BCS-type
interaction. After decoupling the interaction with the help of a
Hubbard-Stratonovich transformation, the resulting action is split into two
parts $S[\mathbf{\Psi },\check{\Delta}]=S_{1}[\mathbf{\Psi },\check{\Delta}%
]+S_{2}[\check{\Delta}]$, where
\begin{eqnarray}
&&S_{1}[\mathbf{\Psi },\check{\Delta}]=\\
&&\quad \int dx\;\mathbf{\Psi }^{\dagger }(x)\left[ i\hat{\tau}_{3}\partial
_{t}-\check{H}(x)+\mu +\check{\Delta}\left( x\right) \right] \mathbf{\Psi }%
(x),\no\\
&&S_{2}[\check{\Delta}]=-\frac{2\nu }{\lambda }\int dx\;\mbox{tr}\left[
\check{\Delta}^{+}\hat{\sigma}_{1}\check{\Delta}\right].  \label{eq:S2}
\end{eqnarray}
Here, $\nu $ is the density of states per one spin projection at the Fermi level
and $\mu $ is the chemical potential. The dimensionless coupling constant in the Cooper channel $\lambda$ is positive for an attractive interaction. Hereafter, we use the hat
symbol as in $\hat{\tau}_{3}$ to denote $2\times 2$ matrices in Keldysh ($K$%
, retarded/advanced) or Gor'kov-Nambu ($N$, particle/hole) spaces. By $\hat{\sigma}_{i}$ and $\hat{\tau}_{i}$ we denote the Pauli matrices in $K$ and $N$ space, correspondingly. The
check symbol as in $\check{H}$ denotes $4\times 4$ matrices in the direct
product space $K\otimes N$. The trace operation $\mbox{tr}$ in Eq.~(\ref%
{eq:S2}) comprises both $K$ and $N$ spaces. The short notation $x=(\mathbf{r}%
,t)$ is used, and the time integration covers the interval $(-\infty,\infty
)$. The single-particle Hamiltonian $\check{H}$ is defined as
\begin{equation}
\check{H}=-\frac{1}{2m}\left( \nabla -ie\mathbf{A}(\mathbf{r})\hat{\tau}%
_{3}\right) ^{2}+U\left( \mathbf{r}\right) +e\varphi (\mathbf{r}),
\end{equation}%
with a static disorder potential $U$, scalar $\varphi $ and
vector potentials $\mathbf{A}$, and electron mass $m$ and charge $e$. In
the action, $\mathbf{\Psi }$ is a four component vector of Grassmann fields
with the following structure:
\begin{equation}
\mathbf{\Psi}=\left(
\begin{array}{c}
\mathbf{\psi }_{1} \\
\mathbf{\psi }_{2}%
\end{array}%
\right)_{K},\quad \mathbf{\psi }_{i}=\left(
\begin{array}{c}
\chi _{i{\uparrow }} \\
\chi _{i\downarrow }^{\ast }%
\end{array}%
\right) _{N}
\end{equation}
\begin{equation}
\mathbf{\Psi }^{\dagger }=\left( \mathbf{\psi }_{1}^{\dagger },\mathbf{%
\psi }_{2}^{\dagger }\right) _{K},\quad \mathbf{\psi }_{i}^{\dagger }=(\chi
_{i\uparrow }^{\ast },-\chi _{i\downarrow })_{N}.
\end{equation}
All terms in the electronic action
\thinspace $S_{1}$ are diagonal in $K$-space except the order parameter
field $\check{\Delta}=\hat{\Delta}_{0}\hat{\sigma}_{0}+\hat{\Delta}_{1}\hat{%
\sigma}_{1}$, where $\hat{\Delta}_{0}$ and $\hat{\Delta}_{1}$ will be referred as classical ($cl$) and quantum ($q$) components of the order parameter. These components are non-diagonal in $N$ space: $\hat{\Delta}%
_{i}=\Delta _{i}\hat{\tau}_{+}-\Delta _{i}^{\ast }\hat{\tau}_{-}$, where $%
\hat{\tau}_{\pm }=\frac{1}{2}\left( \hat{\tau}_{x}\pm i\hat{\tau}_{y}\right)
$. We arrange the classical and quantum order parameter fields into the
vector $\vec{\Delta}=(\Delta _{cl},\Delta _{q})^{T}$.

The electronic Green's function for the system reads:
\begin{eqnarray}
\check{G}\left( x,x^{\prime }\right) =-i\int D\mathbf{\Psi }D\check{\Delta}%
\; \mathbf{\Psi }\left( x\right) \mathbf{\Psi }^{+}\left( x^{\prime }\right) %
\mbox{e}^{iS[\mathbf{\Psi },\check{\Delta}]}.
\end{eqnarray}%
This expression can be cast in the form
\begin{eqnarray}
\check{G}(x,x^{\prime })=\int D\check{\Delta}\;\check{G}_{\Delta
}(x,x^{\prime })\;\mbox{e}^{iS_{GL}[\vec{\Delta}]},
\end{eqnarray}%
where the Ginzburg-Landau action is determined by
\begin{eqnarray}
S_{GL}[\vec{\Delta}]=-i\ln \int D\mathbf{\Psi }\;\mbox{e}^{iS[\mathbf{\Psi }%
,\check{\Delta}]},
\end{eqnarray}%
while
\begin{equation}
\check{G}_{\Delta }(x,x^{\prime })=-i\frac{\int D\mathbf{\Psi }\;\mathbf{%
\Psi }(x)\mathbf{\Psi }^{\dagger }(x^{\prime })\;\mbox{e}^{iS_{1}[\mathbf{%
\Psi },\check{\Delta}]}}{\int D\mathbf{\Psi }\;\mbox{e}^{iS_{1}[\mathbf{\Psi
},\check{\Delta}]}}.
\end{equation}%
This Green's function depends on the specific configuration of the order
parameter field $\check{\Delta}$.

Physical quantities can be obtained in terms of the disorder-averaged $%
\left\langle \check{G}(x,x^{\prime })\right\rangle _{dis}$, which can be found as
\begin{equation}
\left\langle \check{G}(x,x^{\prime })\right\rangle _{dis}=\int D\vec{\Delta}%
\left\langle \check{G}_{\Delta }(x,x^{\prime })\right\rangle _{dis}\;\mbox{e}%
^{i\left\langle S_{GL}[\vec{\Delta}]\right\rangle _{dis}}.  \label{averaging}
\end{equation}
Here, we average the electronic Green's function separately from the bosonic
action. This is a valid approximation for films with dimensionless conductance $g\gg 1$; taking into account cross-correlations between the two terms would give corrections to the Drude conductivity of the order of $1/g^2$, while we are only interested in corrections of the order of $1/g$.

The electric current is related to the Keldysh component of $\left\langle \check{G}(x,x^{\prime })\right\rangle _{dis}$:
\begin{equation}
\mathbf{j=-}\frac{e}{2m}\left( \mathbf{\nabla }_{\mathbf{r}}-\mathbf{\nabla }%
_{\mathbf{r}^{\prime }}\right) \left. \left\langle G^{K}\left( x,x^{\prime
}\right) \right\rangle _{dis}\right. _{x\rightarrow x^{\prime }}-\frac{ne^{2}%
}{m}\mathbf{A},
\label{eq:current0}
\end{equation}
where $n$ stays for the density of electrons.

In the following, it will be convenient to use the quasiclassical approximation \cite{eilenberger68, larkin69, Kopnin01}.
The quasiclassical Green's function can be introduced as follows. First, one
performs the Wigner transform of the disorder-averaged Green's function as
\begin{equation}
\left\langle \check{G}_{\Delta }(\mathbf{p},\mathbf{r},t_{1},t_{2})\right%
\rangle _{dis}=\int d{\bm\rho }\;\mbox{e}^{-i\mathbf{p}{\bm\rho }%
}\;\left\langle \check{G}_{\Delta }(x_{1},x_{2})\right\rangle _{dis},
\end{equation}%
where $\mathbf{r}=\frac{1}{2}(\mathbf{r}_{1}+\mathbf{r}_{2})$, ${\bm\rho }=(%
\mathbf{r}_{1}-\mathbf{r}_{2})$. Next, the quasiclassical Green's function
is obtained by integration over the energy variable $\xi =\frac{p^{2}}{2m}%
-\mu $ which describes the distance from the Fermi surface:
\begin{eqnarray}
&&\check{g}_{\mathbf{n}}(\mathbf{r},t_{1},t_{2})=  \notag \\
&&\quad \frac{i}{\pi }\int_{-\infty }^{\infty }d\xi \;\left\langle \check{G}%
_{\Delta }\left( \mathbf{n}\left( p_{F}+{\xi }/{v_{F}}\right),\mathbf{r}%
,t_{1},t_{2}\right) \right\rangle _{dis}.\;
\end{eqnarray}%
In this equation, $v_{F}$ denotes the Fermi velocity.

In the diffusive limit higher angular harmonics are suppressed and a formulation in
terms of the angular-averaged Green's function is possible:%
\begin{equation}
\check{g}(\mathbf{r},t_{1},t_{2})=\int d\mathbf{n}\;\check{g}_{\mathbf{n}}(%
\mathbf{r},t_{1},t_{2}).
\end{equation}
The function $\check{g}$ satisfies the nonlinear Usadel equation \cite{usadel70,smith86}:
\begin{equation}
D\mathbf{\hat{\nabla}}\left( \check{g}\cdot \mathbf{\hat{\nabla}}\check{g}%
\right) -\left\{ \hat{\tau}_{3}\partial _{t},\check{g}\right\} +i\left[
\check{\Delta}-e\check{\varphi},\check{g}\right] =0,  \label{usadel}
\end{equation}%
where the symbol $\cdot$ is used to denote a convolution in time, i.e., integration in
the intermediate time variable. The spatial derivative has the following structure: $\mathbf{\hat{\nabla}}%
\check{g}=\mathbf{\nabla }\check{g}-ie\left[ \hat{\tau}_{3}\mathbf{A},\check{%
g}\right].$ An important constraint imposed on the quasiclassical Green's function is
that it has to satisfy the normalization condition
\be
(\check{g}\cdot \check{g}%
)(t,t^{\prime })=\check{1}\delta \left( t-t^{\prime }\right).\label{eq:normalization}
\ee

In what follows we are interested in Gaussian fluctuations. This means, that
the film is considered to be not too close to the superconducting
transition. The width of the non-Gaussian region is determined by
the Ginzburg number $Gi$; in the case of disordered films $Gi~\sim g^{-1}$. The precise criterion for the range of
validity of this approximation depends on the quantity in question. Concerning transport phenomena, the non-Gaussian region is wider than for thermodynamics and has been estimated to be of the order of $\sqrt{Gi}$ for the thermal
phase transition \cite{larkin01}, i.e. it covers the temperature regime for which $|T-T_c|\lesssim \sqrt{Gi}T_c$. To the best of our knowledge, there is no such calculation for the quantum transition (a study of the effect of fluctuations on the critical magnetic field exists \cite{galitski03}). In this paper, we assume that we are
always outside the region of non-Gaussian fluctuations.

Let us now turn to the discussion of the Ginzburg-Landau action. As long as
we are interested in Gaussian fluctuations, we need to know $S_{GL}[ \vec{%
\Delta}] $ only up to the second order in $\vec{\Delta}$. Noting the
relation
\begin{eqnarray}
\frac{\delta \left\langle S_{GL}[\vec{\Delta}]\right\rangle _{dis}}{\delta
\Delta _{i}^{\ast }(x_{1})}&=&i\mbox{tr}\left[ \hat{\sigma}_{i}\hat{\tau}%
_{-}\left\langle \check{G}_{\Delta }(x_{1},x_{1})\right\rangle _{dis}\right]
\notag \\
&&-\frac{2\nu }{\lambda }(\hat{\sigma}_{1}\vec{\Delta}(x_{1}))_{i},
\end{eqnarray}%
we can obtain
\begin{eqnarray}
\left\langle S_{GL}[ \vec{\Delta}] \right\rangle _{dis}&=&\int dx_{1}dx_{2}\;%
\vec{\Delta}^{\dagger }(x_{1})  \label{SGL} \\
&&\times \left[ -\frac{2\nu }{\lambda }\hat{\sigma}_{1}\delta _{x_{1},x_{2}}+%
\hat{\Pi}(x_{1},x_{2})\right] \vec{\Delta}(x_{2}),  \notag
\end{eqnarray}%
where
\begin{equation}
\hat{\Pi}_{ij}(x_{1},x_{2})=i\left. \frac{\delta \mbox{tr}\left[ \hat{\sigma}%
_{i}\hat{\tau}_{-}\left\langle \check{G}_{\Delta }(x_{1},x_{1})\right\rangle
_{dis}\right] }{\delta \Delta _{j}(x_{2})}\right\vert _{\vec{\Delta}=0}.
\label{polarization}
\end{equation}
Importantly, the appearing Green's function at coinciding times and space
points is related to the quasiclassical Green's function, and we can write
\begin{eqnarray}
\hat{\Pi}^{ij}(x_1,x_2)=\pi\nu \left.\frac{\delta \mbox{tr}\big[\sigma_i\hat{\tau}%
_-\hat{g}(\mathbf{r}_1,t_1,t_1)\big]}{\delta \Delta_j(x_2)}\right|_{\vec{%
\Delta}=0}.  \label{Pig}
\end{eqnarray}
This result shows that knowledge of the quasiclassical Green's function,
i.e., the solution of the Usadel equation, also allows finding the $GL$
action. This observation considerably simplifies the scheme of calculation of the Gaussian corrections. With the help of the $GL$ action, in turn, one can obtain the order
parameter correlation function, which is needed for the calculation of the
current.

The charge density and electric current are expressed in terms of the angular-averaged
Green's function $\check{g}$ in the following way\cite{smith86}:
\begin{equation}
\rho \left( \mathbf{r},t\right)=-e\nu \left( 2e\phi +\frac{\pi }{2}
\mbox{tr}\left\langle \hat{\sigma}_{1}\hat{g}\left( \mathbf{r},t,t\right)
\right\rangle \right)\label{density}
\end{equation}
and
\begin{equation}
\mathbf{j}\left( r,t\right) = \frac{e\pi \nu D}{2}\mbox{tr}\left\langle
\hat{\tau}_{3}\hat{\sigma}_{1}\mathbf{\check{j}}\left( \mathbf{r},t,t\right)
\right\rangle   \label{current},
\end{equation}
with $\mathbf{\check{j}}=\check{g}\cdot \mathbf{\hat{\nabla}}\check{g}$. The
angular brackets in this equations symbolize averaging with the action $S_{GL}$. Relation (\ref{current}) follows from Eq.~(\ref{eq:current0}) in the diffusion approximation. Aiming for the needed accuracy (the leading order approximation in $g^{-1}$), it is sufficient to determine $\mathbf{\check{j}}$ up to the second order in the fluctuating field $\Delta$ \emph{before} the expansion in the electric field is performed.

\section{Solution of the Usadel equation and the order parameter correlation function}
\label{sec:solution}

For practical calculations, one needs to resolve the normalization
condition (\ref{eq:normalization}) for the quasiclassical Green's function explicitly. In the framework of a mean-field treatment, one works with the
classical order parameter field $\Delta _{cl}$ only. In this case ($\Delta _{q}=0$)
the Green's function can be parameterized as
\begin{equation}
\check{g}=\left(
\begin{array}{cc}
\hat{g}^{R} & \hat{g}^{K} \\
0 & \hat{g}^{A}%
\end{array}%
\right),  \label{gk}
\end{equation}%
with $\hat{g}^{K}=$ $\hat{g}^{R}\cdot \hat{h}-\hat{h}\cdot \hat{g}^{A}$ and $%
(\hat{g}^{R}\cdot \hat{g}^{R})_{t,t^{\prime }}=(\hat{g}^{A}\cdot \hat{g}%
^{A})_{t,t^{\prime }}=\hat{1}\delta_{t-t^{\prime }}.$ However,
in the presence of the quantum order parameter fluctuation (i.e., for finite $\Delta_q$) this structure is broken
and a more general parametrization needs to be considered. In that case, one can generalize (\ref{gk}) to take into account fluctuations up to the second order in $\Delta$:
\begin{equation}
\check{g}=\left(
\begin{array}{ccc}
\hat{g}^{R}-\hat{h}\cdot \hat{g}^{Z} &  & \hat{g}^{R}\cdot \hat{h}-\hat{h}%
\cdot \hat{g}^{A}-\hat{h}\cdot \hat{g}^{Z}\cdot \hat{h}-\hat{g}^{W} \\
\hat{g}^{Z} &  & \hat{g}^{A}+\hat{g}^{Z}\cdot \hat{h}%
\end{array}%
\right).  \label{parametrization}
\end{equation}%
In particular, the lower left corner of this matrix is not equal to zero\cite{zala01,kamenev09}. This parametrization has the following property:
\begin{equation}
\label{props}
\left( \hat{g}^{R}\right) ^{2}=\left( \hat{g}^{A}\right) ^{2}=\hat{1}\delta_{t-t^{\prime }}
+\mathcal{O}\left( \Delta ^{4}\right)
\end{equation}%
The matrix
\begin{equation}
\hat{h}=\left(
\begin{array}{cc}
h_e & 0 \\
0 & h_h
\end{array}
\right)
\end{equation}
is called generalized distribution function \cite{Kopnin01}. Matrices $\hat{g}^{Z,W}$ are diagonal and appear only in the second order in $\Delta$. This holds provided the distribution function $\hat{h}$ satisfies the following normal metal diffusion equation:
\begin{equation}
\label{diffusion}
D\nabla^2\hat h -\left[\partial_t+i e \phi\hat\tau_3,\hat h\right]=0.
\end{equation}

For the purpose of our calculation, we may assume $\hat{g}^{Z}=\hat{g}^W=0$. In the case of $\hat{g}^Z$ the reason is the following. For the calculation of the current, the
Green's function needs to be inserted into the corresponding expression (\ref%
{current}) and subsequently averaged over order
parameter configurations. There can be two kinds of contributions to the
current originating from $\hat{g}^{Z}$. First, if it is not combined with any
other term arising due to fluctuations, it should be averaged by itself. Since the
lower-left corner of the averaged Green's function must equal zero
in the Keldysh formalism $\left\langle \hat{g}^Z \right\rangle=0$, contributions of this first type vanish
automatically. The second kind of contribution appears when combining $\hat{g}^{Z}$
with other terms arising due to fluctuations in formula (\ref{current}). Since $\hat{g}^{Z}$ itself is already quadratic in $\Delta$, this procedure generates contributions to the current which are at least of the fourth order in $\Delta$. These terms are beyond the accuracy of our calculation. The same argument applies to contributions originating from $\hat{g}^W$, only in this case the average $\left\langle \hat{g}^W\right\rangle$ does not vanish identically, but is $\mathcal{O}(\mathbf{E}^2)$, as discussed in Appendix \ref{sec:collision}. Therefore, there is no need to keep $\hat{g}^W$ when studying linear response in the electric field. To conclude, for the purpose of our calculation we may work with the simple parametrization given in Eq.~\ref{gk}.

In what follows, we consider static and
homogeneous electric $\mathbf{E}$ and magnetic $\mathbf{B}$ fields and find
it convenient to work in a gauge with time-independent
electromagnetic potentials: $\mathbf{E=-\nabla }\phi $ and $\mathbf{B}=%
\mbox{curl}\mathbf{A}$ with $\phi =-\mathbf{Er},~\mathbf{A}=\left(
0,Bx,0\right).$ Under these conditions and in the absence of superconducting fluctuations, the retarded and advanced sectors of the quasiclassical Green's function are diagonal in N-space and take a particularly simple form
\begin{equation}
\hat{g}^{R}(t_{1},t_{2})=-\hat{g}^{A}(t_{1},t_{2})=\hat{\tau}_{3}\delta_{t_{1}-t_{2}}.
\label{eq:metallic}
\end{equation}

For a closed system, i.e. in the absence of a connection to an external bath, the distribution function $\hat{h}$ is not yet fixed. Indeed, equation (\ref{diffusion}) has infinitely many solutions. In the presence of interactions, it is convenient to work with the distribution function corresponding to the state of local thermal equilibrium with spatially varying chemical potential:
\begin{equation}
\label{h0}
\hat{h}=\left(
\begin{array}{cc}
h_e & 0 \\
0 & h_h
\end{array}
\right),\quad h_{e,h}=\mathcal{H}(\epsilon\mp e\phi\left(x\right))
\end{equation}
where
\be
\mathcal{H}(\epsilon )=\tanh \frac{\epsilon}{2T}.
\ee
This particular choice is especially convenient for linear response studies, because deviations of $\left<\hat{g}^W\right>$ from zero which arise due to interactions are pushed to the second order in the electric field. This considerably simplifies perturbation theory. Note that temperature is still arbitrary and is determined by the heat balance with a substrate or with contacts. Meanwhile, by neglecting $\left<\hat{g}^W\right>$ we dismiss the heating effect of the electric field.

In the presence of superconducting fluctuations, the quasiclassical Green's function acquires off-diagonal components in $N$-space. For the analysis of the Gaussian fluctuation regime, the deviations from the simple form given in Eq.~\ref{eq:metallic} are small and may be treated as a perturbation. With this in mind, we resolve the remaining constraints (\ref{props}) as:

\begin{gather}
\hat{g}^{R}=\left(
\begin{array}{cc}
1-\frac{1}{2}f\cdot f^{\ast } & f \\
f^{\ast } & -1+\frac{1}{2}f^{\ast }\cdot f%
\end{array}%
\right), \label{gr}\\
\hat{g}^{A}=\left(
\begin{array}{cc}
-1+\frac{1}{2}\bar{f}\cdot \bar{f}^{\ast } & -\bar{f} \\
-\bar{f}^{\ast } & 1-\frac{1}{2}\bar{f}^{\ast }\cdot \bar{f}%
\end{array}%
\right),  \label{ga}
\end{gather}
From the solution of the Usadel equation it will follow that $f$, $\bar{f}$ etc. are $\mathcal{O}(\Delta)$. The functions $f$ and $f^{\ast }$ as well as $\bar{f}$ and $\bar{f}^{\ast }$
are considered as independent: they become complex conjugates of each other only when $\Delta_q$ is neglected.

We introduce parametrization (\ref{parametrization}) into Eq.~(\ref{usadel}) and neglect terms of the third order in $\Delta$. As a result,  we find for $f$ the equation $\mathcal{C}^{-1}f=V$, where the operator $\mathcal{C}^{-1}$ is given by
\be
\mathcal{C}^{-1}=D\mathbf{\hat{\nabla}}^{2}-\partial
_{t_{1}}+\partial _{t_{2}}
\ee
and the gauge invariant derivative is: $\mathbf{\hat{\nabla}}%
f=\left( \mathbf{\nabla }-2ie\mathbf{A}\right) f$. As one may notice, this equation describes the response of the field $f$ to the order
parameter $\Delta$, which enters this equation in the following combination:
\begin{equation}
\label{V}
V_{t_{1},t_{2}}(\mathbf{r})=2i\left[ \Delta _{cl}\left(\mathbf{r},t_{1}\right) \delta
_{t_{1}-t_{2}}+h_{e}\left(\mathbf{r},t_{1}-t_{2}\right) \Delta _{q}\left(\mathbf{r},t_{2}\right)
\right].
\end{equation}
Similar equations arise for $f^*$, $\overline{f}$, and $\overline{f}^*$ with appropriately modified differential operators and functions $V^*$, $\overline{V}$ and $\overline{V}^*$. Taking into account the explicit form of $h_{e,h}$ one may conclude that $\bar{f}_{t_{1},t_{2}}=-f_{t_{2},t_{1}}$ (the same property holds for $f^{\ast}$). Note that a static electric
potential does not enter the equation for $f$. This is one of the advantages of the gauge in which the electric field is expressed through the scalar potential.

The equation for $f$ can easily be solved after a Fourier transformation to the frequency domain according to
\begin{equation}
f\left( t_{1},t_{2}\right) =\int f\left( \epsilon _{1},\epsilon _{2}\right)
e^{-i\left( \epsilon _{1}t_{1}-\epsilon _{2}t_{2}\right) }\left( d\epsilon
_{1}\right) \left( d\epsilon _{2}\right).
\end{equation}%
Here we used notation $(d\epsilon )=d\epsilon /2\pi $. To proceed further, we pass to the Landau level (LL) basis with eigenfunctions $%
\psi_{np}\left( \mathbf{r}\right) $ of the kinetic energy
operator
\begin{equation}
-D(\nabla -2ie\mathbf{A})^{2}\psi_{np}\left( \mathbf{r}\right)
=\epsilon _{n}\psi_{np}\left( \mathbf{r}\right).
\end{equation}
This equation describes a "particle" with a mass equal to $1/2D$. We choose to work in the Landau gauge, for which the eigenfunctions $\psi_{np}$ are numbered by the momentum $p$ and LL number $n$:
\be
\psi_{np}\left( \mathbf{r}\right) =e^{ipy}\chi _{n}\left(x-p l_B^2\right)
\ee
with magnetic length $l_B={1}/{\sqrt{2|e|B}}$ (for a "particle" with charge $2|e|$) and

\be
\chi _{n}\left( x\right)=\frac{1}{\sqrt{l_B}}\frac{e^{-x^{2}/2 l_B^2}}{\pi^{1/4}\sqrt{2^{n}n!}}H_{n}\left( x/l_B\right).
\ee
Note that a description based on the Usadel equation is valid as long as we consider the regime of classically weak magnetic fields, for which $\omega_c=\frac{|e|B}{m}$ satisfies $\omega_c\tau\ll 1$. This means that the quantization of the orbital motion of the quasiparticles can be neglected. In contrast, the LL quantization of the collective modes and Cooperons $\epsilon _{n}=\Omega _{c}\left( \frac{1}{2}+n\right)$ with $\Omega _{c}=4|e|DB$ may still be important in the region of magnetic fields and temperatures we are interested in.

The solution for $f$ is
conveniently written in terms of the Cooperon propagator, which is diagonal
in the chosen basis:
$\left\langle n,p\right\vert \mathcal{C} \left\vert m,p\right\rangle=
\delta_{mn}C_{n}(\epsilon_1+\epsilon_2)$ with
\begin{equation}
C_n\left( \epsilon \right) =\left( i\epsilon -\epsilon _{n}-\tau _{\phi
}^{-1}\right) ^{-1}.
\end{equation}
Here, we introduced the dephasing time $\tau_{\phi}$. The role of $\tau_{\phi}$ is to provide the long-time decay of the Cooperon, which is necessary to render corrections due to single-particle interference processes finite. These processes include weak localization and the anomalous Maki-Thompson correction (an analog of weak antilocalization) that diverge in the absence of a magnetic field for $\tau_{\phi}^{-1}=0$. Dephasing can be provided by magnetic impurities or inelastic processes, i.e. electron-electron or electron-phonon collisions. For low temperatures, electron-electron collisions dominate. Outside the region of strong fluctuations (i.e., in the Gaussian regime), one can consider the dephasing rate as energy independent and equal to the sum of rates due to the Coulomb\cite{altshuler82} and Cooper channels\cite{larkin72,brenig85}. In our study, we do not specify the dominant dephasing mechanism relevant for $\tau_{\phi}$ and consider it as an independent parameter.

The solution of the equation $\mathcal{C}^{-1} f=V$ for $f$ reads:
\begin{equation}
\label{fsol}
f_{np}\left(\epsilon_{1},\epsilon_{2}\right) =C_{n}\left(2\epsilon
\right) \int V_{\epsilon_1,\epsilon_2}(\mathbf{r}^{\prime}) \psi _{np}^{\ast
}\left( \mathbf{r}^{\prime }\right) d\mathbf{r}^{\prime},
\end{equation}
where
\begin{equation}
V_{\epsilon_1,\epsilon_2}(\mathbf{r})=2i\left[ \Delta _{cl}\left(\mathbf{r},\omega \right)
+h_{e}\left( \mathbf{r},\epsilon +\omega /2\right) \Delta
_{q}\left( \mathbf{r},\omega \right) \right]
\end{equation}
with shorthand notation $\epsilon =\left( \epsilon _{1}+\epsilon _{2}\right)
/2$ and $\omega =\epsilon _{1}-\epsilon _{2}$. Analogous
equations hold for $f^{\ast }$, $\bar{f}$ and $\bar{f}^{\ast }$.

Having found approximate solutions for $\hat{g}^{R}$ and $\hat{g}^{A}$, we
turn to the $GL$ action $S_{GL}$. As follows from Eq.~\ref{SGL} in combination with Eq.~\ref{Pig}, it is sufficient to know $\hat{g}^{R\left(
A\right) }$ at the first order in $\Delta $ to determine $S_{GL}$ in
the Gaussian approximation. We write the GL action in the form:
\begin{equation}
\label{eq:sgl}
S_{GL}[\vec{\Delta}] =\int\mbox{tr}\left(2\nu\vec{\Delta}^{+}(-\omega,\mathbf{r})\mathcal{L}^{-1}(\omega,\;\mathbf{r},\;\mathbf{r^\prime})\vec{\Delta}(\omega,\mathbf{r^\prime})\right)
\end{equation}
with
\begin{equation}
\label{Linv}
\mathcal{L}^{-1}=\left(
\begin{array}{cc}
0 & \mathcal{L}_{12}^{-1} \\
\mathcal{L}_{21}^{-1} & \mathcal{L}_{22}^{-1}%
\end{array}
\right).
\end{equation}
Arguments ($\omega,\;\mathbf{r},\;\mathbf{r^\prime}$) of $\mathcal{L}^{-1}$ are omitted in what follows.

A straightforward calculation according to Eq.~(\ref{SGL}) gives:
\begin{equation}
\label{L21}
\mathcal{L}_{21}^{-1} =\sum_{np}\psi_{np}\left(
\mathbf{r}\right)\psi
_{np}^{\ast }\left( \mathbf{r}^{\prime }\right)\left[ \int \frac{\mathcal{H}_{\epsilon -\omega /2+e\phi
\left( \mathbf{r}\right) }d\epsilon }{2\epsilon +i(\epsilon _{n}+\tau_{\phi}^{-1})}-\frac{1}{\lambda }\right].
\end{equation}
The rest of the elements of $\mathcal{L}^{-1}$ are related to this one according to $\mathcal{L}_{12}^{-1}=\left( \mathcal{L}_{21}^{-1}\right)^{+}$ and
\begin{equation}
\label{FDT}
\mathcal{L}_{22}^{-1}
=\mathcal{B}\left(\omega-e\phi\left(\mathbf{r}\right)-e\phi\left(
\mathbf{r}^{\prime }\right)\right)\left[\mathcal{L}_{21}^{-1}-\mathcal{L}_{12}^{-1}\right],
\end{equation}
where the bosonic distribution function is defined as
\be
\mathcal{B}\left(
\omega \right) =\coth(\omega /2T).
\ee
One can see, that the components of
$\mathcal{L}^{-1}$ are not independent. Just as the components of $\mathcal{L}$, they are related by the
generalized fluctuation-dissipation theorem, see Eq.~(\ref{FDT}), valid in the quasi-equilibrium state. Thus, only $\mathcal{L}_{21}^{-1}$ needs to be calculated explicitly. The
evaluation of the $\epsilon$ integral in Eq. (\ref{L21}) is straightforward and yields:
\begin{equation}
\label{l21}
\mathcal{L}_{21}^{-1}=\sum_{np}\psi_{np}(\mathbf{r})\psi _{np}^{\ast }(\mathbf{r}^{\prime })\mathcal{E}_{n}(\omega-2e\phi(\mathbf{r})),
\end{equation}
where
\be
\mathcal{E}_n(\omega)=\ln \frac{T_{c}}{T}+\psi \left( \frac{%
1}{2}\right)-\psi^R(n,\omega)
\ee
and
\begin{equation}
\psi^{R(A)}(n,\omega)=\psi \left(\frac{%
1}{2}+\frac{\epsilon_n +\tau_{\phi}^{-1}\mp i\omega }{4\pi T}\right).
\end{equation}
We have introduced the BCS transition temperature $T_{c}=%
\frac{2\gamma \omega _{D}}{\pi }\exp \left( -\frac{1}{\lambda}\right)$, where $\omega _{D}$ is the Debye frequency and $\gamma\approx 1.78$. The symbol $\psi$ stands for the Digamma function\cite{Abramowitz72}. In deriving asymptotic expressions, we will use the following properties of this function: $\psi^\prime(1/2)={\pi^2}/{2}$ and $\psi(x)\approx\ln x$ for $x\gg 1$.

The line of the superconducting transition on the mean field level is
determined by the equation $\mathcal{E}_{n=0}(\omega=0)=0$. In the absence of dephasing $\tau_{\phi}=\infty$ this gives for
the upper critical field
\be
B_c(T=0)=\frac{\pi T_{c}}{2\gamma D}.
\ee
Let us discuss the effect of dephasing on the transition line. Since the fluctuation propagator depends on the dephasing rate, the transition temperature is shifted due to $\tau_\phi$. Furthermore, since $\tau_\phi$ depends on the magnetic field as well as on the temperature, the presence of $\tau_\phi$ in the fluctuation propagator changes the shape of the transition line as a whole. Dephasing also affects the magnetoconductivity. This effect has been taken into account in the analysis of the experimental data on magnetoconductivity of thin superconducting InO films\cite{brenig86}.

As can be seen from the right-hand side of Eq.~(\ref{l21}), $\mathcal{L}_{21}^{-1}$ is not translation invariant. However, by splitting off a gauge-dependent factor it can be rewritten in the following form:
\begin{equation}
\mathcal{L}_{21}^{-1}\left(t,\mathbf{r},\mathbf{r}^{\prime }\right)=
e^{-iS_g(t,\mathbf{r},\mathbf{r}^{\prime })}
\mathcal{\bar{L}}_{21}^{-1}\left(t,\mathbf{r}-\mathbf{r}^{\prime }\right),
\end{equation}
where $S_g$ is defined as
\be
S_g(t,\mathbf{r},\mathbf{r}^{\prime })&=&e(\phi(\mathbf{r})+\phi(\mathbf{r^\prime}))t\no\\
&&-e\mathbf{\left(A(r)+A(r^{\prime})\right)\left(r-r^{\prime}\right)}
\ee
and $\mathcal{\bar{L}}_{21}^{-1}$ is translational and gauge invariant. We nevertheless prefer to work with the operator $\mathcal{L}^{-1}$ in its original form.

In order to find the order parameter correlation
functions, one has to invert the operator $\mathcal{L}^{-1}$ given by Eq.~(\ref{Linv}) with the
following result:
\begin{equation}
\mathcal{L}=\left(
\begin{array}{cc}
\mathcal{L}^{K} & \mathcal{L}^{R} \\
\mathcal{L}^{A} & 0%
\end{array}%
\right),
\end{equation}%
where
\begin{equation}
\mathcal{L}^{R}=\left( \mathcal{L}_{21}^{-1}\right) ^{-1},~\mathcal{L}%
^{A}=\left( \mathcal{L}_{12}^{-1}\right) ^{-1},~\mathcal{L}^{K}=-\mathcal{L}%
^{R}\mathcal{L}_{22}^{-1}\mathcal{L}^{A}.  \label{LElements}
\end{equation}
The order parameter correlation functions are given by:
\begin{eqnarray}
&&\left\langle \Delta _{cl}\left( \omega \right) \Delta _{c}^{\ast }\left(
-\omega \right) \right\rangle =\frac{i}{2\nu }\mathcal{L}^{K},  \label{LARK}
\\
&&\left\langle \Delta _{cl}\left( \omega \right) \Delta _{q}^{\ast }\left(
-\omega \right) \right\rangle =\frac{i}{2\nu }\mathcal{L}^{R},  \notag \\
&&\left\langle \Delta _{q}\left( \omega \right) \Delta _{cl}^{\ast }\left(
-\omega \right) \right\rangle =\frac{i}{2\nu }\mathcal{L}^{A},  \notag \\
&&\left\langle \Delta _{q}\left( \omega \right) \Delta _{q}^{\ast }\left(
-\omega \right) \right\rangle =0.  \notag
\end{eqnarray}
In equilibrium, $\mathcal{L}_{E\rightarrow 0}^{R\left(
A\right) }\left( \omega \right)\equiv L^{R\left( A\right) }\left( \omega
\right) $ is diagonal in the LL basis, and reads as follows
\be
L^{R}_{n}\left( \omega
\right)=\mathcal{E}_n^{-1}(\omega).
\ee
For the Keldysh propagator this gives, according to Eq.~(\ref{LElements}):
\begin{equation}
\mathcal{L}_{E\rightarrow 0}^{K}\left( \omega \right) =\mathcal{B}\left(
\omega \right) \left( L^{R}\left( \omega \right) -L^{A}\left( \omega \right)
\right) \equiv L^{K}\left( \omega \right).
\end{equation}

While we have already neglected the heating induced by the electric field, we still keep other nonlinear effects. For example, one may consider the decay of fluctuating Cooper pairs due to the acceleration of the paired electrons caused by the electric field. It was considered before on the basis of the phenomenological theory\cite{schmid69,gorkov70,varlamov92,mishonov02} (with only AL process included). At $T\sim T_{c}$ this effect becomes essential at electric fields of the order of $E\sim T_{c}/e\xi_{GL}$ that can be rather small due to the divergence of the coherence length $\xi_{GL}$ at the transition.

In the following calculations all nonlinear effects will be neglected. In the linear response regime, we need to
find propagators at first order in the electric field. Concerning the dependence of $\mathcal{L}$ on spatial arguments, we will consider it as an
operator in the basis of the LLs, the same is assumed regarding the
position operator $\mathbf{r}$. Hence, in the equations below these two
operators do not commute. We linearize $\mathcal{L}_{21}^{-1}$, looking for the first order correction to its equilibrium value. In the equations below we do not indicate the
frequency dependence of propagators, having in mind that all functions have
the argument $\omega $. Taking into account first-order corrections in the
electric field we write
\be
\mathcal{L}_{21}^{-1}=\left( 1+2e\mathbf{Er}%
\partial _{\omega }\right)\mathcal{E}.
\ee
For $\mathcal{L}^{R}$ this gives:
\begin{eqnarray}
\mathcal{L}^{R} &=&L^{R}+\delta L^{R},  \label{LR}\\
\label{deltalr}
\delta L^{R} &=&-2e\mathbf{E}L^{R}\mathbf{r\partial }_{\omega }\mathcal{E}L^{R}
\end{eqnarray}
and $\mathcal{L}^{A}$ can be found by hermitian conjugation. Let us turn to $%
\mathcal{L}^{K}$. In order to find it, we need $\mathcal{L}_{22}^{-1}$ given by
Eq.~(\ref{FDT}):
\begin{equation}
\mathcal{L}_{22}^{-1}=\mathcal{B}\left( \mathcal{L}_{21}^{-1}-\mathcal{L}%
_{12}^{-1}\right) +e\mathbf{E}\partial _{\omega }\mathcal{B}\left\{
(\mathcal{E}-\mathcal{E}^{\ast}),\mathbf{r}\right\},
\end{equation}
where curly brackets denote an anticommutator.
Plugging this expression into Eq.~(\ref{LElements}), we obtain
\begin{eqnarray}
\label{LK}
\mathcal{L}^{K}&=&L^{K}+\delta {L}^{K}\\
\label{deltalk}
\delta L^{K}&=&\mathcal{B}\left( \delta L^{R}-\delta L^{A}\right)\\
\nonumber
&&-e\mathbf{E}\partial _{\omega }\mathcal{B}L^{R}\left\{ (\mathcal{E}-\mathcal{E}^\ast),%
\mathbf{r}\right\} L^{A}.
\end{eqnarray}
Now, the order parameter correlation functions given by Eqs.~(\ref{LARK}) are fully specified, and we can
proceed to the calculation of the electric current.

To summarize, we have collected the basic elements of the
formalism used for the calculation of the fluctuation conductivity in this
paper. Once the quasiclassical Green's function is found as a solution of
the Usadel equation (\ref{usadel}), the
current can be obtained from Eq.~(\ref{current}). Since the quasiclassical Green's
function is a functional of the order parameter configuration, formula (\ref%
{current}) for the current includes an average with respect to the $GL$ action.
This action, in turn, can be found from the quasiclassical Green's function
via Eqs. (\ref{SGL}) and (\ref{Pig}) and thus a closed scheme is established.
As we have already argued, it will be sufficient for our purposes to work with $\hat{g}$ given by Eq.~(\ref{gk}) where $\hat{g}^{R(A)}$ are defined in Eqs.~(\ref{gr}),\;(\ref{ga}), and the distribution function $\hat h$ presented in Eq.~(\ref{h0}).

\section{Calculation of the electric current}
\subsection{Fluctuation corrections: derivation}
\label{sec:calculation}

Before studying the
fluctuation corrections, we first show how to obtain Drude conductivity from
the formalism. Input are the normal-metal solution of the Usadel equation: $\hat{%
g}^{R}=-\hat{g}^{A}=\hat{\tau}_{3}$ and the distribution function in the
presence of the electric field, Eq.~(\ref{h0}). This gives, according to
Eq.~(\ref{current}), the electric current:
\begin{equation}
\mathbf{j}^{(n)}=e\pi \nu D\mbox{tr}\hat{\tau}_{3}\mathbf{\nabla }\hat{h}=2\nu e^{2}D\mathbf{E}.
\end{equation}
This results in the Drude formula $\sigma _{D}=2\nu e^{2}D$.

Now we turn to the calculation of the fluctuation corrections. Starting with expression
(\ref{current}), we substitute for $\hat{g}$ the parametrization (\ref{gk}) and obtain the following contributions
to the current
\begin{equation}
\mathbf{j}=\mathbf{j}^{(n)}+\mathbf{j}^{(dos)}+\mathbf{j}^{(an)}+\mathbf{j}^{(sc)}.
\end{equation}%
Here, all terms besides $\mathbf{j}^{(n)}$ depend on the realization of the superconducting order parameter $\Delta $
and have to be averaged using the order parameter correlation functions (\ref{LARK}). The fluctuation contributions can be written in the following form (hereafter the derivative is with respect to the energy argument):
\begin{eqnarray}
\mathbf{j}^{(dos)} &=&2\pi e^{2} D \mathbf{E}\int \mathcal{H}^{\prime }\left( \epsilon \right)  \delta \nu \left( \epsilon
\right)  \left( d\epsilon
\right),  \label{jdos} \\
\text{ }\mathbf{j}^{(an)} &=&2\pi e^{2} D \mathbf{E}\int  \mathcal{H}^{\prime }\left( \epsilon \right)\vartheta \left(
\epsilon \right) \left(
d\epsilon \right),  \label{jmt} \\
\mathbf{j}^{(sc)} &=&2\pi  e D \int
\mathcal{H}\left( \epsilon \right) \mathbf{j}^{(s)}\left( \epsilon \right) \left( d\epsilon \right).  \label{jal}
\end{eqnarray}%
The quantities which appear in these expressions are defined as follows
\begin{equation}
\delta \nu \left( \epsilon \right) =-\frac{\nu }{8}\left\langle f\cdot
f^{\ast }+f^{\ast }\cdot f+\left( f\leftrightarrow \bar{f}\right)
\right\rangle _{\epsilon,\epsilon },
\end{equation}
\begin{equation}
\vartheta\left( \epsilon \right) =-\frac{\nu}{4}\left\langle \bar{f}\cdot
f^{\ast }+\bar{f}^{\ast }\cdot f\right\rangle _{\epsilon,\epsilon },
\end{equation}
and
\begin{eqnarray}
\label{js1}
j^{(s)}_{\alpha}\left( \epsilon \right) &=&\frac{\nu}{8}\left\langle f\cdot
\hat{\nabla}_{\alpha}f^{\ast }-\hat{\nabla}_{\alpha}f\cdot f^{\ast }\right.\\
\nonumber
&&\left.-\left( f\leftrightarrow \bar{f}\right)-\left(f\leftrightarrow f^{\ast}\right)\right\rangle
_{\epsilon,\epsilon }.
\end{eqnarray}

\begin{figure}[tbp]
\includegraphics[width=180pt]{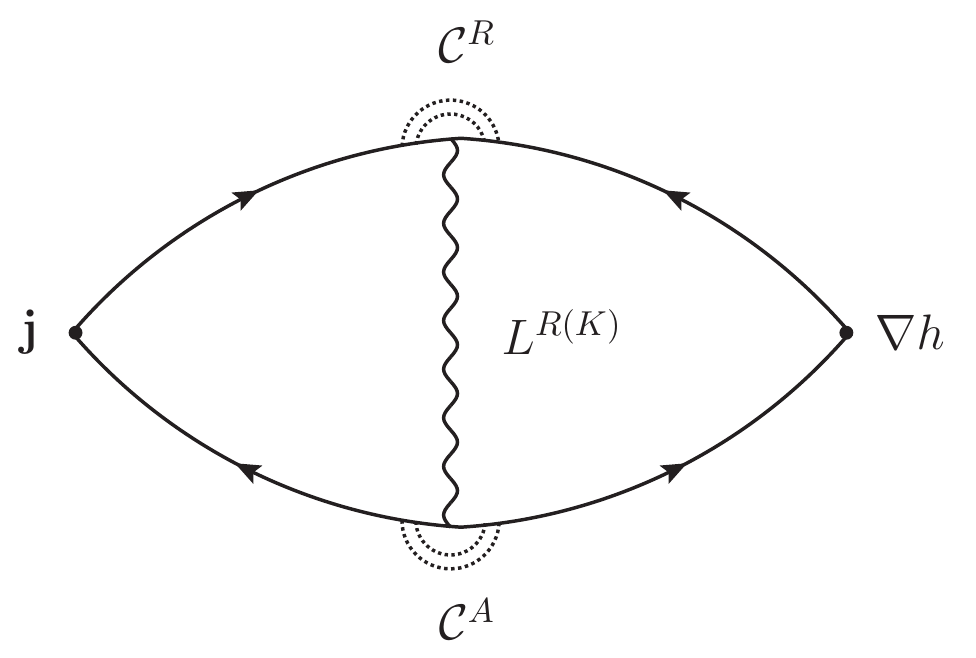}
\caption{Anomalous Maki-Thompson diagram.}
\label{amt}
\end{figure}

The rationale behind this decomposition is the following: (i) The function $\delta\nu(\epsilon)$ describes the correction to
the electronic density of states, see $\delta\nu\left(\epsilon\right)$ in Eq.~(\ref{nu}) below. (ii) The $\vartheta(\epsilon)$-term has a peculiar analytic structure. Indeed, it contains a convolution of $f^*$ and $\bar{f}$, which upon averaging gives rise to a product of two Cooperons of different analytical structure, $\mathcal{C}^R$ and $\mathcal{C}^A$, and the imaginary part of the fluctuation propagator, $\mbox{Im}L^R$. This allows to identify this term with the anomalous Maki-Thompson contribution. For an illustration of this point, we refer to Fig. \ref{amt}.
(iii) The $j^{(s)}_\alpha$-term can be interpreted as the fluctuating supercurrent density. This term is the result of the expansion in the electric field of the fermionic distribution function $h_e$ entering either the combination $V$ (see Eq.~(\ref{V})) or the order parameter correlation function $\mathcal{L}$ (see Eqs.~(\ref{deltalr}) and\;(\ref{deltalk})). The former contribution is purely quantum, while the latter comprises both quantum and classical parts, which are of different importance in the different regions of the phase diagram.

We note that the decomposition (i) - (iii) is very different from the conventional classification based on the diagrams in the Matsubara technique. The difference is related to two main points: a) in the traditional technique a response to a time-dependent vector potential is calculated and b) in the present method there is no need for an analytic continuation.

It is obvious from Eqs.~(\ref{jdos}) and (\ref{jmt}) that $\mathbf{j}^{(dos)}$ and $\mathbf{j}^{(an)}$ contribute only to the longitudinal current, while $\mathbf{j}^{(sc)}$ contributes to the transverse current as well. In this context it should be kept in mind that in the Usadel equation, which was used as a starting point for our calculation, the Lorentz force acting on the quasiparticle was neglected.

To proceed further, we substitute the expressions for $f$, $f^{\ast
}$, $\overline{f}$ and $\overline{f}^{\ast }$ in the LL basis (cf. Eq.~(\ref%
{fsol})) into the expressions above and average them with respect to order parameter fluctuations. The quantities
$\delta\nu(\epsilon)$ and $\vartheta(\epsilon)$ are equilibrium properties of
the system and are independent of the electric field,  and that is why their
calculation is relatively simple. Let us start with the DOS correction which can be understood as a renormalization of the quasiparticle density-of-states:

\begin{eqnarray}
\delta \nu \left( \epsilon \right) &=&\upsilon\sum_n\mbox{Im}\int(d\omega)\; C_{n}^{2}\left(
2\epsilon -\omega \right)\no\\
&&\qquad\times \left[L_{n}^K(\omega)+L_{n}^{R}(\omega)\mathcal{H}\left(\epsilon-\omega\right)\right].\label{nu}
\end{eqnarray}
Here, $\upsilon={1}/{2\pi l_{B}^{2}}$ is the number of states per unit area of a LL. This factor appears with each summation over LLs. In the continuous limit $\upsilon\sum_n\to\sum_q$ and the above expression becomes identical to the one in Eq.~(372) in the review by Kamenev and Levchenko\cite{kamenev09}.
Note that $\int \delta \nu \left( \epsilon \right) d\epsilon =0.$ This is because the interaction cannot change the total number of
single-particle states, but just redistributes them.

Turning to the anomalous MT correction, we find that it is due to a real process. Indeed, $\vartheta(\epsilon)$ can be presented in the following form:
\begin{equation}
\vartheta(\epsilon)=\upsilon\sum_n\frac{\tau_{out,n}^{-1}(\epsilon)}{\epsilon_n+\tau_{\phi}^{-1}},  \label{d}
\end{equation}
where $\tau_{out,n}^{-1}$ is the partial ($n$) out-scattering rate for quasiparticles arising due to the decay of superconducting fluctuations\cite{reizer92}:
\begin{eqnarray}
\tau^{-1}_{out,n}(\epsilon)&=&2\int(d\omega)\;\mbox{Re} C_{n}\left( 2\epsilon -\omega \right)\no\\
&&\qquad \times \mbox{Im}L_{n}^{R}\left(\omega \right) \left[ \mathcal{B}\left( \omega
\right) +\mathcal{H}\left( \epsilon -\omega \right) \right].
\end{eqnarray}
 The discussed correction disappears at zero temperature. This makes it essentially different from the DOS correction which exists down to zero temperature.
The sign of the anomalous MT correction is always positive. It is closely related to weak antilocalization, and can be interpreted as an interference effect in the singlet Cooper channel, enhanced by coherent scattering on the fluctuating order parameter.

Next, we turn to the calculation of the supercurrent $\mathbf{j}^{(s)}$, which is more complicated because non-equilibrium terms in the fluctuation propagators have to be taken into account. The calculation gives:
\begin{eqnarray}
\label{jsx}
j_x^{(s)}\left(\epsilon\right)&=&\frac{eE_x}{8}\upsilon\sum_n\int(d\omega)(n+1)\left\{A_{n,n+1}(\omega,\epsilon)\right\}_-\quad\\
\label{jsy}
j_y^{(s)}\left(\epsilon\right)&=&\frac{eE_x}{8}\upsilon\sum_n\int(d\omega) i(n+1)\no\\
&&\qquad \times \left\{A_{n,n+1}(\omega,\epsilon)-A_{n,n}(\omega,\epsilon)\right\}_+.
\end{eqnarray}
In these equations, the notation $\left\{X\right\}_{\pm}=X\pm\tilde{X}$ is introduced, where $\tilde{X}$ is obtained from $X$ by the substitution $n\leftrightarrow n+1$. The functions $A_{mn}(\epsilon,\omega)$ are defined in the Appendix \ref{sec:supercurrent}.

The next step is to substitute $\delta \nu \left( \epsilon \right),~\vartheta\left( \epsilon
\right) $ and $j^{(s)}_{\alpha}\left( \epsilon \right) $ into the expressions (\ref{jdos})-(\ref{jal}) and to perform the integration in $\epsilon $. The results
of these integrations can be expressed in terms of $\mathcal{E}_n$:
\begin{widetext}
\begin{eqnarray}
\label{DOSFinal}
\delta\sigma^{(dos)}_{\parallel}&=&-2e^{2}D\upsilon\sum_n\int(d\omega)\left[
\mathcal{B}\mbox{Im}\frac{\mathcal{E}_n^{\prime\prime}}{\mathcal{E}_n}+
\mathcal{B}^{\prime}\frac{\mbox{Im}\mathcal{E}_n\mbox{Re}\mathcal{E}_n^{\prime}}{|\mathcal{E}_n|^2}\right],\\
\label{MTFinal}
\delta\sigma^{(an)}_{\parallel}&=&-4e^{2}D\upsilon\sum_n\int(d\omega)\mathcal{B}^{\prime}\frac{\mbox{Im}^2\mathcal{E}_n}{|\mathcal{E}_{n}|^2}
\frac{1}{\tau_\phi^{-1}+\epsilon _{n}},\\
\delta \sigma_{i}^{(sc)}&=&-2e^{2}D\Omega _{c}^{-1}\upsilon \sum_n\int(d\omega)\left(
n+1\right) \left( \mathcal{B} u_{i}+\mathcal{B}^{\prime} v_{i}\right),  \label{ALFinal}
\end{eqnarray}
\end{widetext}
where $i=\parallel,\perp$. For the longitudinal ($\parallel$) conductivity,
\begin{eqnarray}
u_{\parallel}&=&\mbox{Re}\left[ K_{n}K_{n}^{\prime }L_{n}^{R}L_{n+1}^{R}\right],\\
v_{\parallel}&=&2\mbox{Re}K_{n}\mbox{Im}\left[ \mathcal{E}_{n}+\mathcal{E}_{n+1}\right]\mbox{Im}\left[L_{n}^{R}L_{n+1}^{A}\right]\no\\
&&+\mbox{Im}K_{n}\mbox{Im}\left[ L_{n+1}^{R}-L_{n}^{R}
\right]
\end{eqnarray}
with $K_n(\omega)=\psi^R_{n+1}(\omega)-\psi^R_n(\omega).$ For the transversal ($\perp$) conductivity (assuming negatively charged carriers $e<0$ for the rest of the paper; otherwise, the sign of the Hall conductivity should be reversed), we obtain:

\begin{widetext}
\begin{equation}
u_{\perp }=2\mbox{Im}\left[ K_{n}L_{n}^{R}L_{n+1}^{R}\left(\mathcal{E}_{n}^{\prime
}+\mathcal{E}_{n+1}^{\prime }\right) \right]-2\Omega_c \mbox{Re}\left\{ \left(
L_{n}^{R}\right) ^{2}\mathcal{E}_{n}^{\prime }\psi _{n}^{R\prime }\right\}
_{+}-\mbox{Im}\left[ K_{n}^{\prime }\left( L_{n+1}^{R}+L_{n}^{R}\right) %
\right]+\Omega_c \mbox{Re}\left\{ \psi _{n}^{R\prime \prime
}L_{n}^{R}\right\} _{+},
\end{equation}
\begin{equation}
v_{\perp }=-2\mbox{Im}(\psi_n^R+\psi_{n+1}^R)\mbox{Re}K_{n}\mbox{Re}\left[
L_{n}^{R}L_{n+1}^{A}\right]-2\Omega_c \left\{\mbox{Im}\psi _{n}^{R}\mbox{Im}%
\psi _{n}^{R\prime }L_{n}^{R}L_{n}^{A}\right\}_{+}-\mbox{Im}K_{n}\mbox{Re}%
\left( L_{n+1}^{R}+L_{n}^{R}\right)+\Omega_c \mbox{Re}\left\{ L_{n}^{R}\mbox{%
Re}\psi _{n}^{R\prime }\right\}_{+}.
\end{equation}
\end{widetext}

To conclude, we have derived fluctuation conductivity due to electron-electron interactions in the Cooper channel
in the Gaussian approximation. Equations (\ref{DOSFinal})-(\ref{ALFinal}) describe the contribution of superconducting fluctuations to the conductivity everywhere in the $\left(
B,T\right)$ phase diagram (outside the regime of strong fluctuations close
to the transition). In the rest of the paper we discuss different limiting cases and
elaborate on asymptotics of these general formulas.

\subsection{Discussion: Longitudinal conductivity}

\label{sec:results_long}
\begin{figure}[tbp]
\includegraphics[width=265pt]{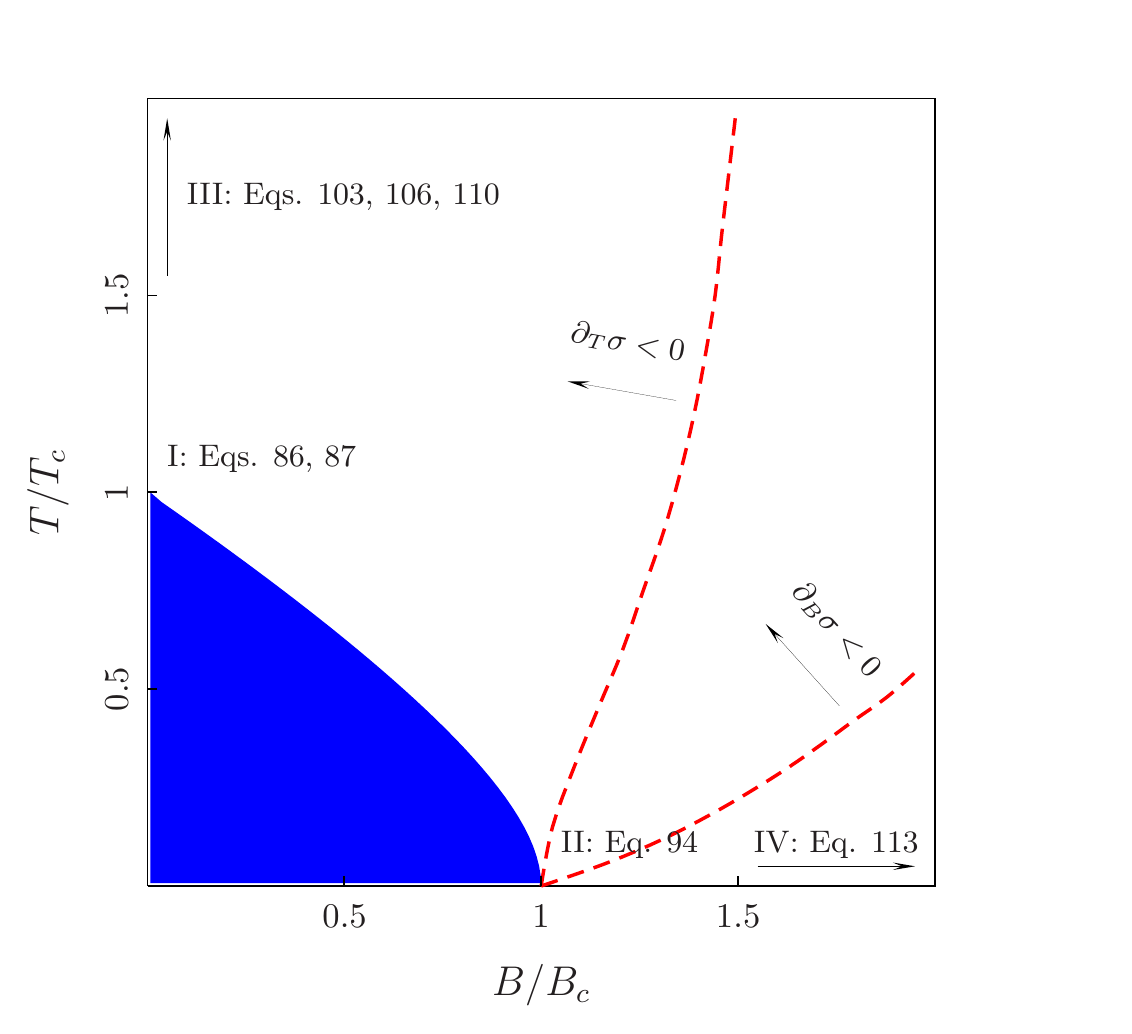}
\newline
\caption{Phase Diagram for the correction to the longitudinal conductivity $\delta\sigma_{xx}$. The corresponding equations are written in the text.}
\label{fig:phasediag}
\end{figure}

At the end of the previous section, we provided general formulas for the fluctuation
corrections to conductivity. In certain asymptotic regions of the phase diagram they are amenable to an analytic treatment. Following this route, we are able to  compare our results to the previous studies. The derived formulas can also be subjected to a numerical analysis, which allows to find the corrections in the entire normal part of the phase diagram.

We will discuss the following asymptotic regions in the phase diagram: The vicinities of the classical (I) and quantum (II) transition points, the region of high temperatures and small
magnetic fields (III) and the region of high magnetic fields and low temperatures (IV). The corresponding regions are indicated on the phase diagram displayed in Fig.~\ref{fig:phasediag}. By means of a numerical evaluation, we locate the line which describes the transition from positive to negative magnetoresistance ($\partial_B\sigma=0$), and the line which characterizes the change of the temperature dependence of the total correction $\partial_T\sigma=0$.

\subsubsection{GL region (I)}
In this region, $\delta\sigma^{(sc)}_{\parallel}$ and $\delta\sigma^{(an)}_{\parallel}$ are the most important. Since the leading contribution comes from small bosonic momenta and frequencies $(\omega,Dq^2\lesssim T-T_c)$, in order to extract the result, one should expand the
equilibrium propagator in $\omega/T$ and $\epsilon_n/T$:
\begin{equation}
\left[ L_{n}^{R\left( A\right) }\left( \omega \right) \right] ^{-1}\approx
\frac{\pi }{8T}\left[ -\tau _{GL}^{-1}-\epsilon _{n}\pm i\omega \right],
\label{LT}
\end{equation}
where
\begin{equation}
\tau_{GL}=\frac{\pi}{8T\ln T/T_c}.
\end{equation}
In this section we assume $\tau_{\phi}\gg\tau_{GL}$ and neglect $\tau_{\phi}$ in the fluctuation propagator.
Substituting the expression for the propagators $L_{n}^{R\left( A\right)}$ to Eqs.~(\ref{MTFinal}) and (\ref{ALFinal}), integrating in frequency (only the term proportional to $\mathcal{B}^{\prime }$ contributes), and performing the summation over the LL index, we obtain:
\begin{equation}
\delta\sigma^{(an)}_{\parallel}=\frac{e^2}{\pi}T\tau _{GL}\left[ \psi \left( \frac{1}{2}%
+s\right) -\psi \left( \frac{1}{2}+s\frac{\tau _{GL}}{\tau_{\phi}}\right)
\right]  \label{mt1}
\end{equation}
and
\begin{equation}
\delta\sigma^{(sc)}_{\parallel}=\frac{2e^2}{\pi }(T\tau _{GL})s\left[ -1-2s\psi \left( s\right)
+2s\psi \left( \frac{1}{2}+s\right) \right],  \label{al1}
\end{equation}
with
\be
s=\left(\Omega_{c}\tau_{GL}\right)^{-1}.
\ee
These results are in agreement with existing calculations. In particular, $\delta\sigma^{(sc)}_{\parallel}$ was obtained phenomenologically by Abrahams et al.\cite{abrahams71}, and the Maki-Thompson contribution was discussed for finite magnetic fields in Ref.~\onlinecite{hikami88}. Note, that the parameter
$s$ divides the region (I) into two parts with a distinct behavior. The zero-field limit is recovered for $s\gg1$:
\begin{equation}
\delta\sigma^{(an)}_{\parallel}=\frac{e^2}{\pi}T\tau _{GL}\ln(\tau_{\phi}/\tau
_{GL}),\;\delta\sigma^{(sc)}_{\parallel}=\frac{e^2}{2\pi}T\tau _{GL}.
\end{equation}
In the absence of a magnetic field, the importance of the anomalous MT correction, $\delta\sigma^{(an)}_{\parallel}$, in comparison with $\delta\sigma^{(sc)}_{\parallel}$ is
determined by the ratio $\tau_{\phi}/\tau _{GL}$. Indeed, the MT term diverges in the absence of dephasing, $\tau
_{\phi }\rightarrow \infty $, and becomes comparable to the AL correction when $\tau _{\phi
}\sim\tau_{GL}$. As the ratio decreases further, the relative importance of the MT correction diminishes.

For completeness, let us discuss the DOS correction in region (I). In the vicinity of the critical temperature, $\delta\sigma^{(dos)}_{\parallel}$ is weakly (only logarithmically) singular. The reason is that interactions preserve the total density of states, and the integration with $\mathcal{H}^{\prime}$ in Eq.~({\ref{jdos}}) is (comparatively) wide: $\epsilon\lesssim T\approx T_c$. For zero magnetic field one gets:
\begin{equation}
\label{dostc}
\delta\sigma_{\parallel}^{(dos)}=-\frac{7\zeta(3) e^2}{\pi^4}\ln T\tau_{GL}.
\end{equation}

A contribution of the same form originates also from the anomalous MT correction as a subleading term, with a numerical coefficient $-14$ instead of $-7$. It is instructive to perform a comparison with the previously known result in this region. For that, one should sum all terms of the kind $\delta\sigma=c\frac{\zeta(3)}{\pi^4}\ln T\tau_{GL}$.
In the diagrammatic calculation,\cite{dorin93} one obtains the coefficient $c=-14$ as the combined contribution of all diagrams with a horizontal interaction line. Those diagrams taken together are often referred to as the DOS-type corrections. In addition, regular MT, AL and anomalous MT diagrams come with the coefficients $c=-7$, $c=14$ and $c=-14$, correspondingly. One can see that only after summation of all logarithmic terms of this kind, the results of the two approaches coincide, and one obtains in both cases a total numerical coefficient $c_{tot}=-21$.

We would like to stress that according to Eq.~(\ref{jdos}) it is the contribution $\delta\sigma_{\parallel}^{(dos)}$ rather than the sum of all horizontal diagrams that should be associated with the suppression of the single-particle density of states.

\subsubsection{Quantum critical point (II)}

In the vicinity of the transition line, for
\begin{equation}
h=(B-B_{c}(T))/B_c \ll 1,
\end{equation}
the most singular contribution comes from the lowest LL, $n=0$. For small temperatures in the vicinity of the Quantum Critical Point (QCP), when
\begin{equation}
t=T/T_{c}\ll 1,
\end{equation} we can simplify the inverse fluctuation propagator using the asymptotic formula for the Digamma function:
\begin{equation}
\mathcal{E}_n(\omega)=-h-\ln \left( 2n+1\right)-\ln
\left( 1-\frac{i\omega }{\epsilon _{n}}\right).
\end{equation}
In this region,  the role of $\tau_{\phi}$ in the fluctuation propagator is mostly to shift the critical magnetic field. We will assume that this shift has already been performed. Besides, it is natural to neglect $\tau_{\phi}$ in the Cooperon, because in the vicinity of the critical point the Cooperon is not singular and $1/\tau_{\phi}$ has to compete with $\Omega_c$.
Substituting the expression for $\mathcal{E}_n(\omega)$ into Eqs.~(\ref{DOSFinal})-(\ref{ALFinal}) and expanding the propagators in $\omega
/\Omega_{c}$, the contributions of all three terms can be written in the form
\begin{equation}
\label{AsymptII}
\delta\sigma_{\parallel}^{(i)}=\frac{e^2}{\pi^{2}}\left[\alpha^{(i)}I_{\alpha }\left( t,h\right) +
\beta^{(i)}I_{\beta }\left( t,h\right) \right]
\end{equation}
with the numerical coefficients
\begin{eqnarray}
\alpha^{(dos)}=-1,~\alpha^{(an)}=0,~\alpha^{(sc)}=\frac{1}{3}, \\
\beta^{(dos)}=-1,~\beta^{(an)}=2,~\beta^{(sc)}=\frac{5}{3}.
\end{eqnarray}
Here
\begin{eqnarray}
\no
I_{\alpha}=\int_0^{\Omega_c}\frac{\omega\mathcal{B}(\omega)d\omega}{\omega^2+(h\Omega_c/2)^2},~
I_{\beta}=-\int_0^{\infty}\frac{\omega^2\mathcal{B}^{\prime}(\omega)d\omega}{\omega^2+(h\Omega_c/2)^2}.
\end{eqnarray}
Evaluating these integrals, we obtain:
\begin{eqnarray}
I_{\alpha}\left( t,h\right) &=&\ln \frac{r}{h}-\frac{1}{2r}-\psi \left(
r\right),~ \\
I_{\beta}\left( t,h\right) &=&r\psi ^{\prime }\left( r\right) -\frac{1}{2r}%
-1
\label{integralsQCP}
\end{eqnarray}%
with
\begin{equation}
r=\frac{1}{2\gamma }\frac{h}{t}.
\end{equation}
Note that when all the contributions are summed up, we get $\alpha=-\frac{2}{3},\;\beta=\frac{8}{3}$, and our result reproduces the one obtained by Galitski and Larkin\cite{galitski01}.

The region of the phase diagram in the vicinity of the $QCP$ can further be subdivided into classical and quantum regions,
depending on the ratio of the parameters $h$ and $t$. The superconducting fluctuations contribute either as classically populated modes or through virtual transitions.
In the quantum region $t\ll h$ the occupation number of the lowest LL of the collective mode is small, and we obtain
\be
\delta\sigma_{\parallel}=-\frac{2e^2}{3\pi^{2}}\ln \frac{1}{h},\qquad (t\ll h).
\ee
In the classical region $t\gg h$, the occupation number is large and the correction changes its character. As a result, it becomes positive:
\be
\delta\sigma_{\parallel}=\frac{2e^2\gamma}{\pi^{2}}\frac{t}{h},\qquad (t\gg h).
\ee

\subsubsection{High temperatures (III) and high magnetic fields (IV)}
In these regions the dominant contributions come from high LLs and, hence, the summation in the LL index can be replaced by an integration. At the same time, the full dependence of the fluctuation propagators on the bosonic frequency should be kept, because the leading contribution comes from a long double logarithmic integration.

Let us first discuss the region (III). We will perform the calculation in the limit of $\ln(T/T_{c})\gg 1$. We start with the analysis of $\delta\sigma^{(dos)}_{\parallel}$. It has a very slow temperature dependence due to
the long integration in energy, which has to be cut off at $\omega,\epsilon
\sim\tau^{-1},$ where the diffusive approximation breaks down. In
view of this fact, only the term proportional to $\mathcal{B}$ (rather then $\mathcal{B}^{\prime }$) gives
the leading contribution, and we can write
\begin{eqnarray}
\delta\sigma^{(dos)}_{\parallel}&=&\frac{e^{2}}{4\pi^{2}}\int\mathcal{B}\left( \omega \right) \mbox{Im}\left[ L^{R}\left( \omega \right)\psi^{R\prime\prime}(\omega)\right]d\omega
d\epsilon\no\\
&=&-\frac{e^{2}}{4\pi ^{2}}\mbox{Im}\int \frac{\mathcal{B}\left( \omega
\right) \partial _{\omega }^{2}\psi \left( \frac{1}{2}+\frac{\epsilon
-i\omega }{4\pi T}\right) d\omega d\epsilon }{\ln T/T_{c}+\psi \left( \frac{%
1}{2}+\frac{\epsilon -i\omega }{4\pi T}\right) -\psi \left( \frac{1}{2}%
\right) }.\no\\
\end{eqnarray}
This integral is logarithmically divergent. As a result, we obtain:
\begin{equation}  \label{DOSHigh}
\delta\sigma^{dos}_{\parallel}= -\frac{e^{2}}{2\pi ^{2}}%
\ln \frac{\ln 1/T_{c}\tau }{\ln T/T_{c}}.
\end{equation}
This correction is similar to the Altshuler-Aronov corrections, but with a scale-dependent coupling constant. This result was first derived by Altshuler et al. \cite{altshuler83}. At very large temperatures ($\ln T/T_c\gg1$) this term dominates the total correction. In the case of a repulsive interaction, it becomes\cite{fukuyama85} $\frac{e^2}{2\pi^2}\ln\ln\frac{1}{T\tau}$.

Let us turn to $\delta\sigma^{(sc)}_{\parallel}$. The term proportional to $\mathcal{B}^{\prime }$ is again small, $\mathcal{O}\left(\ln^{-2}(T/T_{c})\right)$. Another term, which is proportional to $\mathcal{B},$ is more important:
\begin{equation}
\delta\sigma^{(sc)}_{\parallel}=e^2\int_{0}^{\infty }\frac{izdz}{256\pi^{5}}\int_{-\infty
}^{\infty }\frac{dy\coth \frac{y}{2}\psi ^{\prime }\left(\varepsilon\right) \psi ^{\prime \prime }\left( \varepsilon\right) }{\left[ \ln T/T_{c}+\psi \left(\varepsilon\right) \right] ^{2}}
\end{equation}
where $\varepsilon=\frac{1}{2}+\frac{z-iy}{4\pi}$. We first calculate the $y$ integral neglecting $y$ in the denominator. Since only $y\gtrsim 1$ contribute to the leading term, we can substitute $\coth\frac{y}{2}\to \sign y$. This leads to
\begin{equation}
\delta\sigma^{(sc)}_{\parallel}=\frac{e^2}{64\pi ^{4}}\int_{0}^{\infty }\frac{zdz\left[ \psi
^{\prime }\left( \frac{1}{2}+\frac{z}{4\pi }\right) \right] ^{2}}{\left[
\ln T/T_{c}+\psi \left( \frac{1}{2}+\frac{z}{4\pi }\right) \right]
^{2}}.
\end{equation}
The remaining integral comes from $1\lesssim z$ and can be calculated to
give:
\begin{equation}  \label{ALHigh}
\delta\sigma^{(sc)}_{\parallel}=\frac{e^2}{4\pi^{2}}\frac{1}{\ln T/T_{c}}.
\end{equation}
We note, however, that the same term originates from the subleading contribution to $\delta\sigma_{\parallel}^{(dos)}$, but with a different numerical coefficient $\frac{\ln2-1}{2\pi^2}$. Thus, different contributions of the kind $\mathcal{O}(\ln^{-1}T/T_c)$ do not cancel each other.

Let us now turn to $\delta\sigma^{(an)}_{\parallel}$. In the continuous limit, $\upsilon\sum_n\to\sum_q$, Eq.~ (\ref{MTFinal}) reproduces the known result\cite{aslamazov80}. In the limit of $\ln T/T_c\gg 1$, it
can be further simplified to:
\begin{equation}
\label{mthight}
\delta\sigma^{(an)}_{\parallel} =-\frac{e^2}{16\pi ^{2}}\frac{1}{\ln ^{2}T/T_{c}}%
\int_{0}^{\infty }\frac{M(z) dz}{z+1/\left( T\tau _{\phi }\right)}
\end{equation}
with
\begin{equation}
M(z)=\int_{-\infty }^{\infty }\frac{dy\left[ \psi \left( \frac{1}{2}+%
\frac{z-iy}{4\pi }\right) -\psi \left( \frac{1}{2}+\frac{z+iy}{4\pi }\right) %
\right] ^{2}}{\sinh ^{2}\left( y/2\right)}.
\end{equation}
Although this term is formally $\mathcal{O}(\ln^{-2}T/T_c)$, it can still be essential due to the logarithmic divergence at small momenta as it can be seen from Eq.~(\ref{mthight}).  With logarithmic accuracy, we can calculate it as follows:
\begin{equation}
\delta\sigma^{(an)}_{\parallel} =-\frac{e^2}{16\pi ^{2}}\frac{1}{\ln ^{2}T/T_{c}}%
\int_{0}^{1}\frac{M(0)dz}{z+1/\left( T\tau _{\phi }\right)}.
\end{equation}
As a result, we get:
\begin{equation}
\delta\sigma^{(an)}_{\parallel}=\frac{e^2}{12}\frac{\ln T\tau _{\phi }}{\ln ^{2}T/T_{c}}.
\label{MTHigh}
\end{equation}
One should keep in mind, however, that $\tau_{\phi}$ itself depends on $T$. In this region, the anomalous Maki-Thompson correction was considered by several authors, who all obtained the same functional form but with different numerical coefficients\cite{altshuler83,glatz11long,reizer92}. We believe this discrepancy is due to different approximations used for the calculation of $M(0)$.

For high magnetic fields (region (IV)), the situation is to some extent analogous to region (III) with the main difference that the anomalous MT term does not contribute as it is suppressed at small temperature. The dominant corrections originate from $\delta\sigma^{(sc)}$ and $\delta\sigma^{(dos)}$,  and the leading contributions are those which are proportional to $\mathcal{B}\approx \sign \omega$. To proceed, we write the equilibrium propagator in its zero-temperature form:
\begin{eqnarray}
L^{R\left( A\right) }=\ln^{-1}\left(\frac{\Omega_c/2h}{\epsilon _{n}\mp
i\omega}\right),\quad (T\rightarrow 0).
\end{eqnarray}
After the frequency integration, we find that $\delta\sigma_{\parallel}^{(dos)}$ takes the following form:
\begin{equation}
\delta\sigma^{(dos)}_{\parallel}=\frac{e^2}{\pi^2}h\sum_{n}\mbox{li}\left(\frac{1}{h(2n+1)}\right)  \label{DOSHighB}
\end{equation}
with the logarithmic integral function $\mbox{li}(z)=\int_0^z dt/\ln t$. This sum is logarithmically divergent at the upper limit and has to be cut off when the diffusion approximation breaks down, that is at $n\sim N_{max}\gg 1$ with $N_{max}=\frac{1}{hT_c\tau}$. Under these conditions, the sum is dominated by large $n$ and can be found to equal
\begin{equation}
\label{ALHighB}
\delta\sigma^{(dos)}_{\parallel}=-\frac{e^{2}}{2\pi ^{2}}%
\ln \frac{\ln 1/\tau T_{c}}{\ln B/B_{c}}.
\end{equation}
This concludes our discussion of the regions (I-IV) in the phase diagram; the corresponding asymptotic expressions are referenced in Fig. \ref{fig:phasediag}.

\begin{figure}[tbp]
\includegraphics[width=250pt,angle=0]{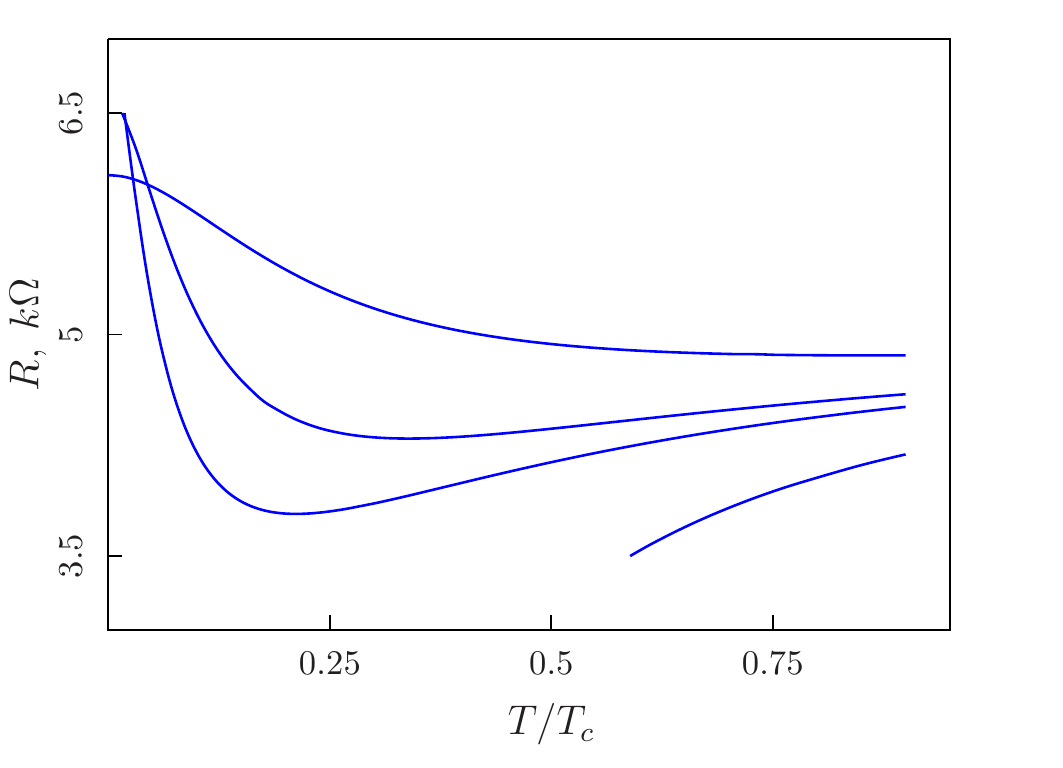}
\caption{Resistance as a function of temperature for magnetic fields $B/B_c=0.9,1.05,1.1,1.3$. The sample parameters are $R_D = 5k\Omega$ and $T_c\tau=10^{-2}$.}
\label{fig:sigmat}
\end{figure}

The results we obtained differ from those given in Ref.~\onlinecite{glatz11long}. This follows from a comparison of the asymptotic behavior in several regions. The most drastic difference, however, concerns the temperature dependence of the resistance for magnetic fields $B>B_c$. The authors of Ref.~\onlinecite{glatz11long} claimed that for small temperatures $T \ll T_c$ the resistance first increases with increasing $T$ and starts to diminish at $T/T_c \gtrsim (B-B_c)/B_c$. As follows from our asymptotic expressions presented in Eqs.~(\ref{AsymptII}) and from the result of the numerical calculation shown in Figs.~\ref{fig:phasediag} and \ref{fig:sigmat}, the situation is opposite. At a fixed magnetic field, the resistance decreases as the temperature increases from zero until the line $\partial_T\sigma=0$ is crossed. Then the resistance starts to grow.

\subsection{Discussion: Hall conductivity}
\label{sec:hall}
We proceed with the discussion of the results for the transverse conductivity presented in Eq.~(\ref{ALFinal}). These expressions represent only those contributions to $\delta \sigma_{\perp}$, which describe a deflection of the supercurrent. In principle other contributions exist, in which quasiparticles are deflected in the transverse direction by the Lorentz force. These contributions are not included in the approximation we apply here. The terms not accounted for by Eq.~(\ref{ALFinal}) include the contribution due to the anomalous MT process, discussed by Fukuyama et al.\cite{fukuyama71} and the contribution $\delta\sigma_{\perp}^{(dos)}$, recently discovered diagrammatically by Michaeli et al.\cite{michaeli12}, which is reminiscent of the density of state suppression. They are related to the corresponding corrections to the longitudinal conductivity as follows:
\begin{gather}
\label{ant}
\delta\sigma_{\perp}^{(an)}=-2\omega_c\tau\delta\sigma^{(an)}_{\parallel},\\
\label{dost}
\delta\sigma^{(dos)}_{\perp}=-\frac{\omega_c\tau}{2}\delta\sigma_{\parallel}^{(dos)}.
\end{gather}
Note, that $\delta\sigma^{(an)}_{\perp}$ and $\delta \sigma^{(an)}_{\parallel}$ cancel each other in the expression for the Hall resistivity $\rho_{xy}=-\sigma_{xy}/(\sigma_{xx}^2+\sigma_{xy}^2)\approx-\sigma_{xy}/\sigma_{xx}^2.$ In contrast, the DOS-corrections give a finite contribution to $\rho_{xy}$.

Let us discuss the contribution to Hall conductivity that arises due to the deflection of the fluctuating supercurrent.
In order to calculate it, it is enough to modify the superconducting fluctuation propagator according to~\cite{aronov95}
\begin{equation}
\label{lsubs}
L_{R(A)}^{-1}(\omega)\to L_{R(A)}^{-1}(\omega)-\varsigma\omega.
\end{equation}
As a consequence of the additional term, the superconducting propagators lose their particle/hole symmetry, i.e., the relation $L^A(-\omega)=L^R(\omega)$ no longer holds. In the framework of the BCS theory, the asymmetry parameter $\varsigma$ can be related to the energy dependence of the density of states at the Fermi level: $\varsigma=-\frac{1}{2\lambda}\frac{d\ln\nu}{d\mu}$ or, equivalently\cite{aronov95}, to the variation of $T_c$ with the chemical potential: $\varsigma=-\frac12\frac{d\ln T_c}{d\mu}$. In the simple model of 3D electrons with a quadratic spectrum, one has $\nu(\epsilon)\approx\nu_0(1+\epsilon/2\epsilon_F)$ and $\varsigma=-1/(4\epsilon_F\lambda)$. For $\lambda\ll1$ the contributions arising from $\delta\sigma^{(sc)}_{\perp}$ are parametrically larger than those arising from $\delta\sigma^{(dos)}_{\perp}$ and $\delta\sigma^{(an)}_{\perp}$. In our calculation of the Hall conductivity, we work in the framework of the quasiclassical approach, using, however, Eq.~($\ref{lsubs}$) for the propagators $L_{R(A)}$. This is a consistent procedure that allows to obtain all contributions to the transverse current proportional to the large parameter $1/\lambda$.

In region (I) after expansion in $\Omega_c (n+1/2)/4\pi T$ and $\omega/4\pi T$ the correction $\delta\sigma_{\perp}^{(sc)}$ takes the form
\begin{equation}
\label{HallGL}
\delta\sigma_{\perp}^{(sc)}=-\frac{16e^2\varsigma\Omega_{c}\left( T\tau _{GL}\right) ^{2}}{\pi ^{2}}f\left( s\right),
\end{equation}
where
\begin{equation}
\nonumber
f\left( s\right) =s^{2}\left[ 1+\psi \left( \frac{1}{2}+s\right) -\psi
\left( 1+s\right) -s\psi ^{\prime }\left( 1+s\right) \right].  \label{HallI}
\end{equation}
In this region, the Hall effect can be considered phenomenologically: the same expression (\ref{HallGL}) was obtained by Aronov and Rapoport\cite{aronov92} (with a different coefficient, it has later been corrected by Aronov et al.\cite{aronov95}) on the basis of the time dependent Ginzburg-Landau theory. For $s\gg 1$, when quantization of the LLs for the superconducting fluctuations is negligible, the expression (\ref{HallGL}) becomes:\cite{fukuyama71}
\begin{equation}
\delta\sigma_{\perp}^{(sc)}=\frac{e^2\varsigma \Omega _{c}}{96}\left( \frac{T}{T-T_{c}}\right)
^{2}.
\end{equation}
The region of applicability of the Eq.~(\ref{HallGL}) is in fact very narrow, and already for $T\gtrsim 1.01 T_c$ one should not expand the full expression for $\delta\sigma_{\perp}^{(sc)}$ in $\Omega_c (n+1/2)/4\pi T$ to get an accurate result. The corresponding formula has been given in Ref. \onlinecite{michaeli12}:
\begin{equation}
\label{eq:ALTc}
\delta\sigma_{\perp}^{(sc)}=\frac{2e^2\varsigma T}{\pi}\sum_{n}\left(  n+1\right)
\frac{\left[L_{n+1}^R\left(0\right)-L_{n}^R\left(0\right)  \right]  ^{3}%
}{\left[  L_{n+1}^R\left(0\right)+L_{n}^R\left(  0\right)  \right]  ^{2}}.
\end{equation}

In region (II), we can limit ourselves to the lowest LL and follow the same route as in the calculation of the longitudinal conductivity. This gives for the quantum regime:
\be
\delta\sigma_{\perp}^{(sc)}=-\frac{e^2\varsigma \Omega _{c}}{3\pi ^{2}}\ln \frac{1}{h},
\ee
and for the classical regime:
\be
\delta\sigma_{\perp}^{(sc)}=\frac{2e^2}{\pi}\frac{\varsigma T}{h}.
\ee
Note, that in this region $\delta\sigma_{\perp}^{(dos)}$ and $\delta\sigma_{\perp}^{(an)}$ exhibit the same singular behavior as $\delta\sigma_{\perp}^{(sc)}$. We do not provide the corresponding expressions, since they follow straightforwardly from Eqs. (\ref{ant}) and (\ref{dost}), together with Eq.~(\ref{AsymptII}).

A more detailed discussion of the corrections to the Hall conductivity due to
superconducting fluctuations is presented in a separate publication, Ref.~\onlinecite{michaeli12}.
\section{Conclusion}
\begin{figure}[tbp]
\includegraphics[width=220pt,angle=0]{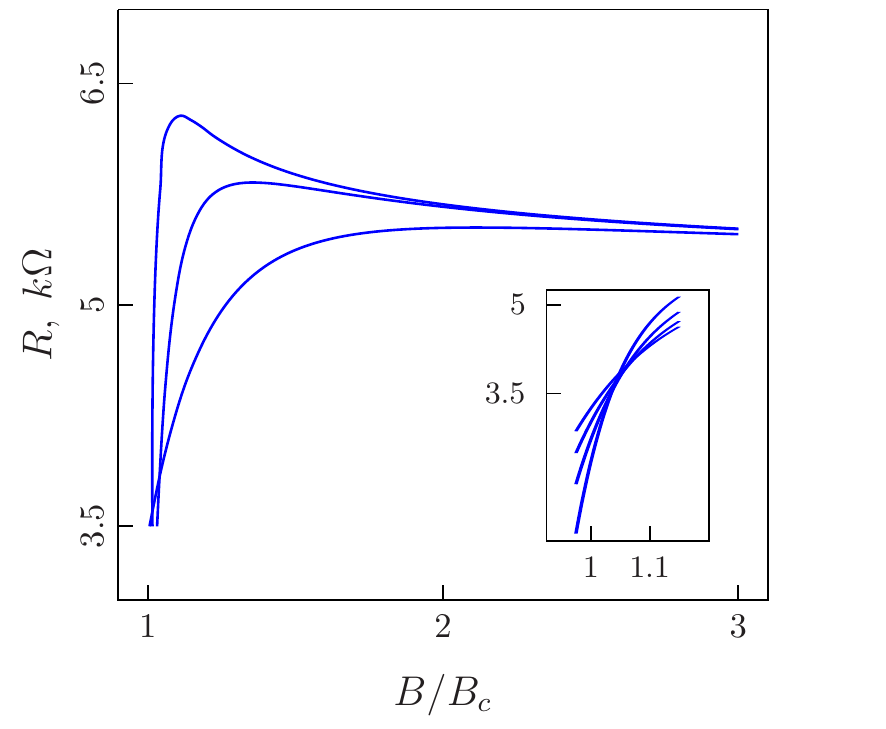}
\caption{Resistance as a function of magnetic field for temperatures $T/T_c=0.03,\;0.1,\;0.35$. Inset: the zoomed region of the approximate crossing for $T/T_c=0.15-0.3$. The sample parameters are $R_D = 5k\Omega$ and $T_c\tau=10^{-2}$.}
\label{fig:sigmab}
\end{figure}
We considered homogeneously disordered films above the superconducting transition
$T>T_c(B)$ and calculated corrections to longitudinal as well as transversal conductivities. Our results are presented by equations (\ref{DOSFinal})-(\ref{ALFinal}).
We analyzed the asymptotic behavior of these corrections in different regions of the phase diagram and provided a comparison with previously published results.

Our results for the Hall effect have recently been used in the description of experimental data by Breznay et. al\cite{breznay10}.
The results for the longitudinal conductivity, Eqs. (\ref{DOSFinal})-(\ref{ALFinal}), can also be useful for the analysis of experiments. They allow for a complete numerical evaluation of the fluctuation corrections to conductivity without any additional approximation, e.g., the lowest Landau level approximation. Exemplary results are presented in Figs.~\ref{fig:sigmat},~\ref{fig:sigmab} for the resistivity $R=(R_D^{-1}+\delta\sigma)^{-1}$ as a function of magnetic field and temperature. A similar behavior of the resistance was observed in the experiment of Baturina et al.\cite{baturina05}. In Ref. \onlinecite{baturina05}, the authors presented a fit to the measured data that was based on the asymptotic expressions (\ref{AsymptII}) derived in Ref.~\onlinecite{galitski01} and reproduced in our work based on a different method. We note, however, that although these expressions provide a good approximation in the vicinity of the QCP, their region of validity does not extend up to the relatively large temperatures and magnetic fields that were considered in the experiment (up to $0.35T_c$ and up to $5B_c$, correspondingly). When fitting this data, the
more precise Eqs.~(\ref{DOSFinal})-(\ref{ALFinal}) should, therefore, be used.

According to the results presented in this work, the resistance curves drawn as a function of the magnetic field exhibit an approximate crossing point for a finite interval of temperatures, as demonstrated in Fig.~\ref{fig:sigmat}. As can be seen from this picture, the curves do not literally cross in a single point, but deviations from this ideal behavior are small. The existence of this approximate crossing point is a consequence of a relatively wide minimum in the $R(T)$ curve for $B=1.05B_c$ as shown in Fig.~\ref{fig:sigmat}. This type of behavior has been observed in several systems; see e.g. Fig. 4 in Ref.~\onlinecite{hebard92}. However, in these experiments the curves continue to cross even at the smallest temperatures, while we did not find this kind of behavior from the Gaussian corrections to conductivity. This could be related to the fact that for such low temperatures the proximity to the QCP becomes of crucial importance, and the present theory is not sufficient because 1) it does not account for the effect of non-Gaussian fluctuations and 2) does not take into account the smearing of the transition by disorder\cite{ikeda02,galitski02}, which is usually observed in this region (see Fig.~2 in Ref.~\onlinecite{hebard84} as an example).

To conclude, we have developed an approach to the calculation of fluctuation conductivity based on the Usadel equation and valid for both the classical as well as the quantum fluctuation regime for arbitrary magnetic fields.
This approach is more physically transparent than conventional perturbation theory based on the Kubo formula and provides a bridge between the phenomenological theory and microscopics. We believe that it may find applications in studies of fluctuation effects out of equilibrium or in hybrid superconductor/normal metal structures.

\begin{acknowledgments}
The authors acknowledge discussions with Karen Michaeli, Brian Tarasinski, Michail Feigel'man, Igor' Burmistrov, Igor' Gornyi, Alex Kamenev, Chandra Varma, and Peter W\"olfle.
GS acknowledges financial support by the Alexander von Humboldt foundation. KT and AMF are supported by the National Science Foundation grant NSF-DMR-1006752 and NHRAP.
\end{acknowledgments}

\begin{appendix}
\section{Collision integrals}
\label{sec:collision}
In this appendix we discuss the quantum components $\hat{g}^{Z,W}$ of the Green's function $\hat{g}^R$.
We parameterize them as
\begin{gather}
\hat{g}^{Z} =\left(
\begin{array}{cc}
z_{1} & 0 \\
0 & z_{2}%
\end{array}%
\right),\;\;\hat{g}^{W} =\left(
\begin{array}{cc}
w_{1} & 0 \\
0 & w_{2}%
\end{array}%
\right)
\end{gather}
and get the following equations:
\begin{equation}
\mathcal{D}^{-1}w_{i}=I^{W}_i,\;
\mathcal{\bar{D}}^{-1}z_{i}=I^Z_i
\end{equation}
with
\begin{equation}
\mathcal{D}^{-1}= D\mathbf{\hat{\nabla}}^{2}-\partial_{t_{1}}-\partial_{t_{2}},\;\mathcal{\bar{D}}^{-1}= D\mathbf{\hat{\nabla}}^{2}+\partial_{t_{1}}+\partial_{t_{2}}.
\end{equation}
The collision integrals $I^{Z}_{1,2}$ are given by
\begin{gather}
\nonumber
I^Z_1=i\left(\Delta_{q}\cdot f^{\ast}-\bar{f}\cdot\Delta_{q}^{\ast}\right),\\
I^Z_2=i\left(\Delta_{q}^{\ast}\cdot f-\bar{f}^{\ast}\cdot\Delta_{q}\right)
\end{gather}
and collision integrals $I^W_{i}=I^W_{i,coll}-I^W_{i,neq}$ by (this separation is motivated below)
\begin{gather}
\nonumber
I^W_{1,coll}=i(f\cdot J_{1}-\bar{J}_1\cdot\bar{f}^{\ast}),\\
\nonumber
I^W_{2,coll}=i(f^{\ast}\cdot J_{2}-\bar{J}_2\cdot\bar{f}),\\
\nonumber
I^W_{1,neq}=2  j_e  \cdot  z_1  \cdot  j_e + j_e  \cdot \bar{f} \cdot  \bar{f}^{\ast\prime}+ f  \cdot  j_h  \cdot  \bar{f}^{\ast\prime}+ \\
\nonumber
f^{\prime} \cdot  j_h  \cdot  \bar{f}^{\ast} + f^{\prime} \cdot f^{\ast} \cdot  j_e,\\
\nonumber
I^W_{2,neq}=2  j_h  \cdot  z_2  \cdot  j_h + j_h  \cdot  \bar{f}^{\ast}  \cdot  \bar{f}^{\prime}+f^{\ast} \cdot  j_e  \cdot  \bar{f}^{\prime}+\\
 f^{\ast\prime} \cdot  j_e  \cdot \bar{f}+ f^{\ast\prime} \cdot  f  \cdot  j_h.
\end{gather}

For convenience, we defined ($j_{e,h}=\pm\nabla h_{e,h}$):
\begin{eqnarray}
\nonumber
J_1&=&\Delta_q^{\ast}-\Delta_c^{\ast} \cdot  h_{e} + h_{h}  \cdot \Delta_c^{\ast}- h_{h}  \cdot \Delta_q^{\ast} \cdot  h_{e},\\
\nonumber
\bar{J}_1&=&\Delta_q-\Delta_c \cdot  h_{h} + h_{e}  \cdot \Delta_c- h_{e}  \cdot \Delta_q \cdot  h_{h}, \\
\nonumber
J_2&=&\Delta_q-\Delta_c \cdot  h_{h} + h_{e}  \cdot \Delta_c- h_{e}  \cdot \Delta_q \cdot  h_{h},\\
\bar{J}_2&=&\Delta_q^{\ast}-\Delta_c^{\ast} \cdot  h_{e} + h_{h}  \cdot \Delta_c^{\ast}- h_{h}  \cdot \Delta_q^{\ast} \cdot  h_{e}.
\end{eqnarray}

While $\left<I^Z\right>=0$ due to causality\cite{kamenev09}, the collision integral $I^W$ does not vanish identically after averaging. Nevertheless, its expansion in the electric field can be shown to start from $\mathbf{E}^2$. First, we note that $I^{W}_{i,neq}$ should be related to the production of the heat. Indeed, $\left<I^{W}_{i,neq}\right>$ is proportional to the Drude result for the electric current $j_{e,h}$. Next, observe that the terms in $\left<I^{W}_{i,neq}\right>$ which are only linear in $j_{e,h}$ are further multiplied by averages which include the spatial gradients of $f$ and vanish in the absence of an electric field, when the system is isotropic. Hence, $\left<I^{W}_{i,neq}\right>=\mathcal{O}(\mathbf{E}^2).$ There is still another term, $I^W_{i,coll}$. For $\mathbf{E}=0$ it corresponds to the collision integral due to Cooper interactions, which enters the kinetic equation and was calculated by Reizer\cite{reizer92}. Let us just note, that if the only source of non-homogeneity is a spatially varying electric potential (as it is in our case), then the collision integral, written in terms of the gauge invariant particle/hole energies should be independent of the spatial coordinates. As such, it cannot depend on the electric field itself, which is a vector, but only on $\mathbf{E}^2$. This is summarized by the equation: $I_{1,2,coll}^W=I_{coll}(\mathbf{E}^2,\epsilon\mp e\phi(x))$. Since for $\mathbf{E}=0$ it vanishes (provided the electronic distribution function $\mathcal{H}$ is thermal) and depends only on $\mathbf{E}^2$, it should be disregarded for the calculations in the linear response.
\section{Calculation of the supercurrent}
\label{sec:supercurrent}
Here we present more details of the calculation of $j^{(s)}_{\alpha}(\epsilon)$. We start with expression (\ref{js1}). After substituting the solution for $f$ and averaging in $\Delta$ we get:
\begin{equation}
\label{jseapp}
j^{(s)}_{\alpha}\left(\epsilon\right)=\frac{1}{8}eE\sum_{mn}\int(d\omega)
I_{\alpha,mn} A_{mn}(\omega,\epsilon)
\end{equation}
Here $I_{\alpha,mn}$ represents the result of integration in the momentum quantum number:
\begin{equation}
\label{dp}
I_{\alpha,mn}=2i\int(dp)
\mbox{Im}\left(\psi_{mp}(\mathbf{r})\hat{\nabla}_{\alpha}\psi_{np}^{\ast}(\mathbf{r})\right)
\left\langle np\right\vert x \left\vert mp\right\rangle
\end{equation}
and $A_{mn}=\sum_k A_{mn}^{(k)}$ has several contributions, which arise from different ways to expand propagators
or bosonic/fermionic distribution functions in the electric field. The next step is to calculate integral (\ref{dp}): taking into account
$\left\langle n,p\right\vert x \left\vert m,p\right\rangle=x_{nm}+p l_B^2\delta_{nm}$,
where $x_{nm}$ are matrix elements, calculated with $\chi_{n}(x)$, we obtain:
\begin{equation}
I_x(m,n)=2\upsilon x_{nm}\partial_{mn},
\end{equation}
\begin{equation}
I_y(m,n)=\frac{2i}{l_B^2}\upsilon(x_{nm}x_{mn}-\delta_{mn}(x^2)_{mn}).
\end{equation}

We also take into account:
\begin{equation}
x_{mn}=\frac{l_B}{\sqrt{2}}(\sqrt{n+1}\delta_{m,n+1}+\sqrt{n}\delta_{m,n-1})
\end{equation}
\begin{equation}
\partial_{mn}=-\frac{1}{\sqrt{2}l_B}(\sqrt{n+1}\delta_{m,n+1}-\sqrt{n}\delta_{m,n-1})
\end{equation}

 and obtain the result, presented in (\ref{jsx}), (\ref{jsy}).

\end{appendix}

\bibliographystyle{apsrev}
\bibliography{usadel}

\end{document}